\documentclass[a4paper,10pt]{article}
\usepackage{jheppub} 
\usepackage{lineno}

\newcommand{\pd}{\partial}
\newcommand{\cpd}{\nabla}

\newcommand{\const}{\mathrm{const}}

\newcommand{\Kc}{\mathcal{K}}
\newcommand{\Mc}{\mathcal{M}}
\newcommand{\Fc}{\mathcal{F}}

\newcommand{\Hc}{\mathcal{H}}
\newcommand{\pr}{^{\prime}}
\newcommand{\prt}{^{\prime\prime}}

\arxivnumber{26xx.xxxx} 

\title{\boldmath Is there ghost and tachyon free bounce\\in UV complete gravity theory?}

\author[a]{Hao Hu,}
\author[a,b]{Alexey S. Koshelev,}
\author[a]{and Abhishek Naskar}
\affiliation[a]{Center for Fundamental Physics, School of Physical Science and Technology,\\ ShanghaiTech University,
393 Middle Huaxia Road, Shanghai 201210, China}
\affiliation[b]{Departamento de F\'isica, Centro de Matem\'atica e Aplica\c{c}oes (CMA-UBI), Universidade da Beira Interior, 6200 Covilh\~a, Portugal }

\emailAdd{huhao12023@shanghaitech.edu.cn, askoshelev@shanghaitech.edu.cn,\\
naskara@shanghaitech.edu.cn}

\abstract{Analytic infinite derivative gravity theories provide a renormalizable and ghost-free description of gravity around covariantly constant backgrounds. These theories can have non-singular bouncing Universe solutions. In this paper we aim to address a question whether it is possible to realize a bouncing solution without the presence of a ghost or a tachyon instability in this framework.
We perform a detailed analysis of degrees of freedom in $(1+3)$ formalism around Minkowski and de Sitter space-times. As a result it becomes clear that on a very general basis one cannot construct an instability free bounce without a negative cosmological constant. An analysis of known bouncing solutions in this model shows that an analyticity of higher derivative form factors in combination with solutions parameters result in the presence of a ghost radiation.
Being motivated by the idea of resolving the cosmological singularity problem we proceed by analyzing scalar and tensor modes anyway. Scalar modes appear to not influence Cosmic Microwave Background observations at all, while tensor modes spectrum is computed and the corresponding implications are discussed.}

\makeatletter
\gdef\@fpheader{}
\makeatother

\begin{document}
\maketitle
\flushbottom
\section{Introduction}
Stelle's theory of gravity \cite{Stelle:1976gc} known to contain a ghost in the spectrum albeit being renormalizable can be extended by means of infinite derivative operators (known as form factors) to a gravity theory which is renormalizable and free of Ostrogradski ghosts  \cite{Krasnikov:1987yj,Kuzmin:1989sp,Tomboulis:1997gg,Modesto:2011kw,Modesto:2014lga,Modesto:2015ozb,Biswas:2011ar,Biswas:2016egy}. Moreover, papers \cite{Modesto:2013oma,Koshelev:2025pxg} conclude that such a setup can result in a finite gravity theory. The form factors appear to be entire functions of the d'Alembertian operator. Consequently, as functions they are analytic at zero of their argument maintaining thereby properly the IR limit of Einstein's GR. The necessity of exactly infinite number of derivatives is dictated by the demand that in order to avoid new poles in the propagator the only appropriate combination is an exponent of an entire function of the d'Alembertian. Thus the non-locality {implied by the infinite derivatives,} is the price for removing perturbative ghosts.
This framework is known as Analytic Infinite Derivative (AID) gravity (and more generally AID theories).

To avoid ghosts around a specific background and simultaneously satisfy the restrictions obtained from the QFT considerations, the form factors have to be chosen appropriately \cite{Tomboulis:1997gg,BasiBeneito:2022wux,Biswas:2016egy}. The weak point of this approach is that in fact form factors can be adjusted around only a single chosen maximally symmetric space-time (MSS) while other backgrounds, even MSS, would be spoiled by alien degrees of freedom (DOF-s) often with curious complex masses \cite{Koshelev:2007fi}. Such emerging modes are discussed in the literature while have not acquired a full understanding \cite{Koshelev:2020fok}. An interesting similarity with respect to the presence of complex mass states can be traced to the papers on Lee-Wick gravity and fakeon models \cite{Lee:1969fy,Lee:1970iw,Anselmi:2019ukt}.
This infinite-derivative construction of gravity can  naturally accommodate cosmological solutions. In particular in several papers (see \cite{Koshelev:2020xby,KoshelevAAS} and refs. therein) $R^2$-like or Starobinsky inflationary solutions ($R$ being the Ricci scalar) are investigated within this framework.\footnote{Inflationary solutions can also be realized within infinite derivative scalar field theories \cite{Barnaby:2007yb,Koshelev:2020fok,Koshelev:2025hdj}.}
In particular it was shown in these papers that infinite-derivative gravity setup leads to distinctive features in scalar non-Gaussianity and can also lead to blue-tilted tensor perturbations. Cosmic inflation in this approach does not suffer from ghost instabilities.
On a general note AID gravity closely resembles models which arise in the Asymptotic safety approach \cite{Dou:1997fg,Percacci:2007sz,Reuter:1996cp,Reuter:2001ag}, especially in the form factor formulation of the asymptotically safe gravity \cite{Knorr:2019atm}.

Being more specific, the aforementioned form factors enter the model via quadratic in curvature terms forming structures as $R\Fc_R(\Box)R$,  $R_{\mu\nu}\Fc_R(\Box)R^{\mu\nu}$ and $C_{\mu\nu\alpha\beta}\Fc_C(\Box)C^{\mu\nu\alpha\beta}$, or their linear combinations. Here $\Box$ is the covariant d'Alembertian, $R_{\mu\nu}$ is the  Ricci tensor,  and $C_{\mu\nu\alpha\beta}$ is the Weyl tensor. Given that form factors are entire functions they can be represented as the Taylor series, say at zero, as follows $\Fc(\Box)=\sum\limits_n f_n\Box^n/\Mc^{2n}$ and one can naturally arrange terms in the powers of the d'Alembertian. Coefficients $f_n$ are supposed to be real as they appear as model parameters in the Lagrangian. The scale $\Mc$ is introduced to make all the coefficients $f_n$ dimensionless. It is important to note that while for one and the same mass dimension other terms can be written the just suggested terms form a special subclass. Indeed it is the result of \cite{Biswas:2016egy} that these and only these terms may give a contribution to a propagator around a MSS. That is, any other term will either give no contribution to a propagator or give a contribution similar to already mentioned terms. This leads to the observation that the presented terms will be dominant at high energies because they are the terms having the maximal possible number of derivatives acting at once on a metric tensor for a given mass dimension. In other words upon quantization the corresponding metric perturbation will be multiplied by the highest possible degree of momenta. This implies that even if other terms are added to the model, those which are quadratic in curvature and have higher derivative form factors must produce a leading contribution to the behavior of any solutions which involve a very high energy stage or regime (like an initial stage of the Universe evolution or a black hole interior).

The above explanation suggests that advertised additional terms in a gravity action become more and more relevant in the very beginning of the Universe. Going further, one can naturally consider a {Bouncing Universe} as a replacement for the Big Bang with an important wishful expectation to resolve the cosmological singularity  \cite{Gasperini:1992em,Enqvist:2001zp,Cardoso:2008gz,Novello:2008ra,Brandenberger:2012zb,Battefeld:2014uga,Brandenberger:2016vhg,Raveendran:2018yyh,Ijjas:2018qbo}. In a bouncing scenario the Universe undergoes a contraction phase which at some moment in its evolution violates the Null Energy Condition and goes through an actual bounce, and then ends up in the expansion phase. The bounce phase is obviously a very high energy stage where gravity modifications must be important and most likely dominating the Universe dynamics. This strongly motivates us to analyze bouncing scenarios in the framework of AID gravity.

Throughout the years several models are proposed for non-singular bouncing Universe using higher-derivative gravity \cite{Nariai:1971sv,Abramo:2009qk,Cai:2015emx}, $f(R)$ gravity \cite{Bamba:2013fha,Nojiri:2014zqa,Nojiri:2019lqw,Singh:2018xjv}, ekpyrotic scenarios \cite{Khoury:2001wf,Khoury:2001zk,Lehners:2007ac,Buchbinder:2007ad,Levy:2015awa,Brandenberger:2020eyf}, matter bounce scenarios \cite{Brandenberger:2009yt,Qiu:2010ch,Cai:2011tc,Brandenberger:2012zb}, Horndeski/Gallileon theories \cite{Easson:2011zy,Cai:2012va,Qiu:2015nha,Koehn:2015vvy,Ijjas:2016tpn,Akama:2017jsa,Mironov:2018oec,Banerjee:2018svi,Ageeva:2022asq}. In all of these descriptions, appearance of Ostrogradski ghosts or gradient instabilities are shown to be present. Especially there are no-go theorems stating that within Horndeski/Gallileon theories non-singular solutions suffer from gradient or ghost instabilities \cite{Kobayashi:2016xpl,Libanov:2016kfc,Akama:2017jsa}. However these no-go theorems can be evaded by considering beyond Horndeski or degenerate higher order scalar tensor (DHOST) theories, but in the latter models issue of super-luminality arises \cite{Creminelli:2016zwa,Kolevatov:2017voe,Ye:2019sth,Ilyas:2020qja,Zhu:2021whu,An:2025xeb}.

AID gravity theories natively accommodate exact analytic cosmological bouncing solutions \cite{Biswas:2005qr,Biswas:2012bp,Kolar:2021qox,Kumar:2021mgc,Kumar:2020xsl,Koshelev:2012qn}.  The AID constructions of gravity do not belong to the family of Horndeski/Gallileon theories and thus do not necessarily obey the no-go theorems mentioned above. Moreover, given the powerful framework to tackle the issue of Ostrogradski ghosts, it was expected that a non-singular, ghost-free description of a bouncing Universe can be realized. In \cite{Biswas:2005qr,Biswas:2012bp} a class of exact bouncing solutions is explored in details within AID gravity framework. The analyzed exact bouncing solution requires the presence of a positive cosmological constant and radiation. It was proposed that the problem of ghosts can be bypassed if the radiation is not a ghost.

With the present paper we want to stress that the issue of ghost in a bouncing cosmology within AID gravity theories {is a subtle one and} was never studied in full. In fact two aspects were not explored in details:
\begin{itemize}
\item
The first aspect is that previous studies were not discussing comprehensively the characteristics and behavior of a scalar DOF which appears for a bouncing solution. The reason that a scalar appears at all is that at least known bouncing solutions which were constructed and analyzed are based on an ansatz $\Box R=r_1 R+r_2$ where $r_{1,2}$ are constants, both non-zero. Non-zero $r_2$ results in a non-zero cosmological constant and in some cases solution asymptotes to the corresponding (Anti-)de Sitter configuration (which is a MSS) \cite{Biswas:2005qr,Biswas:2012bp}. It will be shown as part of our current study that the condition $r_1\neq0$ in combination with a ghost-free condition around the MSS asymptotic (where the conditions in question are derived in \cite{Biswas:2016egy}) inevitably leads to an appearance of a massive scalar with the mass exactly $\mu^2=r_1$. The outcome of the subsequent analysis in this paper is that two requirements of making this scalar stable, i.e. neither a ghost nor a tachyon, and radiation not a ghost simultaneously is in contradiction. In simple words this bouncing scenario is not free of unstable modes.
In \cite{Koshelev:2012qn} another solution satisfying the above mentioned ansatz was found which does not require radiation but results in super-inflation (i.e.~the Hubble parameter grows with the cosmic time) and its cosmological relevance is not clear.
Going further, one of the result in \cite{Koshelev2023solo} (and {refs. therein}) tells that with just a traceless matter cosmological solutions beyond the above ansatz are almost impossible unless the space-time does not allow for a d'Alembertian operator whose eigen-functions form a complete basis in the $L_2$ space. Assuming the latter situation as a marginal one we proceed with using the ansatz if no matter with a non-vanishing trace is included.
\item
The second aspect is that a possible term $C_{\mu\nu\alpha\beta}\Fc_C(\Box)C^{\mu\nu\alpha\beta}$ was not considered in previous papers on bounce in AID gravity. This term is obviously zero on cosmological solutions since the Weyl tensor is zero on conformally flat space-times but it is the cornerstone ingredient for building a renormalizable gravity as it is the one changing the UV behavior of the graviton propagator \cite{Modesto:2013oma,Koshelev:2025pxg}. Naively it can affect perturbations including scalar perturbations. We will take care of this term and finally demonstrate that in fact scalar perturbations are not touched by it. Nevertheless it will be shown that its presence requires extra constraints on the model parameters in order to avoid ghosts in the tensor sector and to keep form factors as entire functions --- a property needed to work with them using a Taylor series representation.
\end{itemize}

A possible presence of ghosts or tachyons is an alarming sign. However, a perspective to resolve the cosmological singularity is a strong motivation to continue pursuing the study of bouncing scenarios, in particular in AID gravity given that according to our arguments above one cannot bypass higher-derivative corrections in consistent gravity models (see also \cite{Ruhdorfer:2019qmk} for an EFT of gravity study). In fact, living with ghosts can be not that impossible as advocated in \cite{Hawking:2001yt} (see also \cite{Koshelev:2020fok,Koshelev:2025hdj} showing that often ghost-like DOF are classically quickly decaying modes, as well as a very recent paper which considers ghosts from the perspective of the Krein space quantization \cite{Bateman:2026eyj}, and related papers \cite{Buoninfante:2026mve,Buoninfante:2026jfj}). Tachyons even though unstable can become normal particles in the proper vacuum of the model. While finding such a mechanism is not the topic of the present paper, scenarios of this kind are widely used in many physically relevant theories.

Even with the issues of ghost or gradient instabilities (or issues of super-luminality) within finite derivative models of bounce, considerable efforts were put to show that  scalar and tensor perturbations can have  observationally relevant spectral tilt and amplitude under certain circumstances \cite{Cai:2008qw,Allen:2004vz,Buchbinder:2007ad,Raveendran:2018why,Bozza:2022bmb}.
AID gravity theories provide a ghost-free renormalizable description of gravity around Minkowski space-time which is why it can be thought as a natural modification to GR. Also as discussed earlier at the high energy scales of early phases of the Universe the higher derivative effects would give the leading order contribution. So it is natural to study bouncing scenarios in this framework  especially given that it provides non singular bouncing solutions.
While we do not dismiss the instability issues we think that a deeper study is required to understand whether the presence of ghosts or tachyons can be fully circumvented.

We thus will compute the scalar and tensor mode functions around a couple of bouncing solutions which have de Sitter asymptotics, describing their general characteristics and  leading order effects.  Around the de Sitter limit of these solutions the form factors allow one massive scalar DOF and two usual massless tensor DOF, and in the current scenario only these modes characterize the scalar and tensor perturbations of the Universe. The reason why the radiation (normal or ghost) is not relevant here is as follows: even if a radiation is present, after the bounce when the  Universe transits to the de Sitter expansion phase the radiation will quickly dilute away and hence at late times we can neglect its contribution to the scalar perturbations. Also, around the  bounce the form factors being adjusted to produce no new excitations around the de Sitter asymptotic, will introduce extra DOF-s for both scalar and tensor sectors which appear with a correct sign of kinetic terms but inevitably with complex masses squared. These modes can become stable, i.e. classically decaying, and do not affect the Cosmic Microwave Background (CMB) observables or the background if the bounce phase is short and if the real and imaginary parts of the masses satisfy the condition $\mathrm{Im} (m^2)^2 < 9 H_{{dS}}^2 \mathrm{Re}(m^2)$ \cite{Koshelev:2020fok,Tokareva:2024sct,Koshelev:2025hdj} for the Hubble parameter around the de Sitter asymptotic. Appearance of this relation is subject to the shape of form factors which is under our control as the form factors are model parameters. With these considerations we will argue that observable scalar perturbations can not be produced within the current framework due to the fact that present scalar modes are massive. For the tensor modes we will show how  the form factors affect the spectrum by solving the tensor equation of motion through the contracting, bouncing and expanding phases of the Universe.

The paper is organized as follows. In Section~\ref{sec:AIDintro} we review the AID gravity construction and discuss the exact cosmological bouncing solutions known in this theory. In Section~\ref{sec:Mink} we analyze the theory around the Minkowski space-time; establish all possible ghost-free form factors around this background using the $(1+3)$ formalism for perturbations; and analyze whether these form factors are compatible with the solution generating ansatz. In Section~\ref{sec:ScalarPert} we repeat this analysis for a more complicate situation of the de Sitter space-time. In Section~\ref{ghostfreeyesno} we find new important constraints required to guarantee a ghost-free configuration in this model around bouncing solutions which asymptote to a de Sitter space-time. We also show that a cosmologically least motivated configuration with a super-inflation behavior tends to avoid instabilities around the bounce phase. In Section~\ref{sec:spectra}
we analyze the observational implications of scalar and tensor perturbations for bouncing scenarios in AID gravity model. We conclude by summarizing our results in Section~\ref{sec:conclusion}. Also a couple of technical Appendices are included to introduce notations and technical parts of the computations.

\section{AID Gravity and exact cosmological bounce solutions}\label{sec:AIDintro}
We consider the following action of AID gravity
\begin{eqnarray}\label{eq:masterL}
    {S} = \int d^4 x \sqrt{-g} \Bigg[\frac{M_P^2}{2}R + \frac{\lambda}{2} R \Fc_R (\Box) R + \frac{\lambda}{2} C_{\mu\nu\alpha\beta} \Fc_C (\Box) C^{\mu\nu\alpha\beta} - \Lambda +\mathcal{L}_M \Bigg].
\end{eqnarray}
We work in 4 dimensions with the metric signature $(-+++)$, $M_P$ is the Planck mass, $R$ is the Ricci scalar, $C_{\mu\nu\alpha\beta}$ is the Weyl tensor, $\Lambda$ is the cosmological constant, $\Fc_R (\Box)$ and $\Fc_C(\Box)$ are form factors built of the covariant d'Alembertian $\Box=g^{\mu\nu}\nabla_\mu\nabla_\nu$ with $\nabla_\mu$ being a covariant derivative, and $\lambda$ is an arbitrary constant which is used to control collectively the effect of all the terms beyond GR. The matter part is described by the Lagrangian $\mathcal{L}_M$. The form factors are entire functions of $\Box$ operator and can be expanded around $\Box = 0$ in Taylor series as
\begin{equation}
    \Fc_R(\Box) = \sum_{n=0}^{\infty} f_n \frac{\Box^n}{\Mc^{2n}},
    ~\Fc_C(\Box) = \sum_{n=0}^{\infty} f_n^{C} \frac{\Box^n}{\Mc^{2n}},
\end{equation}
where $\Mc$ is the non-locality scale,
and $f_n$ and $f_n^C$ are the corresponding Taylor series coefficients which are real constants. Reality of these coefficients is naturally expected as they play role of the model parameters. They can be treated as EFT coefficients in the derivative expansion. We will put $\Mc=1$ for a while and will restore it only at the stage of discussing physics of interesting solutions. At this stage the Ostrogradski statement is the only indication that stopping the Taylor series at some finite order will lead to ghosts. A general ghost free construction around MSS \cite{Biswas:2016egy} indeed demonstrates that an infinite tower of derivative is needed to circumvent this problem. Also, we do not write terms containing the Ricci tensor or any second rank tensor because they can be replaced equivalently using the already present terms as long as only the propagator around MSS is concerned \cite{KoshelevAAS}.

In this Section we will continue in a sketchy manner as most of the most presented here facts were derived and discussed in the already cited papers. We mainly will emphasize the most crucial points which are to be used in the subsequent analysis.

Equations of motion corresponding to action \eqref{eq:masterL} can be written as
\begin{eqnarray} \nonumber
    \Big[M_P^2 + 2 \lambda \Fc_R(\Box) R\Big] G_{\nu}^{\mu} &=& T_{\nu}^{\mu} - \Lambda ~ \delta_{\nu}^{\mu} + \lambda ~ \mathcal{K}_{\nu}^{\mu} -\frac{\lambda}{2} \delta_{\nu}^{\mu}\Big(\mathcal{K}_{\sigma}^{\sigma} + \mathcal{K}_1\Big)  \\ \nonumber
    &-& \frac{\lambda}{2} \delta_{\nu}^{\mu} R \Fc_R(\Box) R
    + 2\lambda \Big(g^{\mu \rho} \nabla_{\rho} \partial_{\nu} - \delta_{\nu}^{\mu} \Box\Big) \Fc_R(\Box) R \\ \label{eq:eom}
    &+& 2 \lambda\Big(R_{\alpha \beta} + 2\nabla_{\alpha} \nabla_{\beta}\Big) \Fc_C (\Box) C_{\nu}^{~\alpha \beta \mu}+O(C^2),
\end{eqnarray}
here $G_{\nu}^{\mu}$ is the Einstein tensor, $T_{\nu}^{\mu}$ is the energy-momentum tensor corresponding to $\mathcal{L}_M$, $R_{\alpha \beta}$ is the Ricci tensor, and $\mathcal{K}_{\nu}^{\mu}$ and $\mathcal{K}_1$ have the following form
\begin{eqnarray}
    \mathcal{K}_{\nu}^{\mu} = \sum_{n=1}^{\infty} f_n \sum_{l = 0}^{n-1} \partial_{\nu}\Box^l R ~ \partial^{\mu} \Box^{n-l-1}R, ~
    \mathcal{K}_1 = \sum_{n=1}^{\infty} f_n \sum_{l = 0}^{n-1} \Box^l R ~ \Box^{n-l}R.
\end{eqnarray}
It is instructive to write down the trace equation which is
\begin{equation}
  E=M_P^{2}R-4\Lambda
-6\lambda\Box\Fc_R(\Box) R
-\lambda(\Kc_\sigma^\sigma +2\Kc_{1})+T+O(C^2)=0.
\label{EOMtrace}
\end{equation}
To find solutions to the full system of equations (\ref{eq:eom}) one can start by solving the latter trace equation. Cosmological FLRW type metrics {as a solution to} equation \eqref{EOMtrace} with a traceless matter such that $T=0$ can be found using the following ansatz that was initially proposed in \cite{Biswas:2005qr,Biswas:2012bp}
\begin{equation}\label{eq:ansatz}
    \Box R = r_1 R + r_2,
\end{equation}
where both $r_{1,2}$ are constants. A solution to the trace equation with a traceless matter will fix the system up to perhaps a radiation fluid (which is the only possible traceless perfect fluid). Its energy can be obtained by substituting the obtained metric into the $(00)$ equation. Note that it was proven in \cite{Koshelev2023solo} that under very general assumptions solutions will obey this ansatz, and until now no FLRW type solutions beyond this ansatz for a traceless matter are known (modulo very simple configurations which can be obtained without imposing any ansatz, like MSS backgrounds, or a situation when $R$ is a constant while it is not a MSS space-time).
Skipping the details which can be found in just cited papers the following algebraic conditions arise if one wants a solution to ansatz (\ref{eq:ansatz}) to be a solution to trace equation (\ref{EOMtrace})
\begin{eqnarray}\label{eq:ansatzImp}
    \Fc_{R}^{(1)}(r_1) = 0; ~~ \frac{r_2}{r_1}  (\Fc_{1} - f_0)= - \frac{M_P^2}{2\lambda} + 3 r_1 \Fc_{1}; ~~ \Lambda = - \frac{M_P^2}{4} \frac{r_2}{r_1}.
\end{eqnarray}
where $\Fc_{R}^{(1)}(r_1)$ is the derivative with respect to the argument and $\Fc_{1}=\Fc_{R}(r_1)$. Here it is implicitly assumed that $r_1\neq0$ since $r_1=0$ is a special and physically irrelevant scenario effectively corresponding to a zero Plank Mass \cite{Kumar:2020xsl}.
It is obvious form the last set of constraints that a non-zero $\Lambda$ demands $r_2 \neq 0$ and vice versa. In particular, when $r_2=0$ the Starobinsky inflation can be exactly embedded within this model \cite{Koshelev:2016xqb}.\footnote{With $r_2 = 0$ the ansatz becomes $\Box R = r_1 R$ which is the trace equation of a local $R^2$ model of Starobinksy inflation.}


To discuss cosmological solutions we introduce a spatially flat FLRW metric
\begin{equation}
    ds^2 = -dt^2 + a(t)^2 d\vec x^2 = a(\tau)^2 \big(-d \tau^2 + d\vec x^2\big),
\end{equation}
with all the notations and conventions (mostly absolutely standard) accumulated in Appendix~\ref{appNotations}.
Known solutions of interest which represent bouncing scenarios are as follows:
\begin{enumerate}
    \item
    Bouncing type I \cite{Biswas:2005qr,Biswas:2012bp}:
    \begin{equation}\label{eq:bouSolcosh}
         a(t) = a_0 \cosh(\sigma t),~H=\sigma\tanh(\sigma t),~R=6\sigma^2(1+\tanh(\sigma t)^2).
    \end{equation}
    This solution satisfies ansatz (\ref{eq:ansatz}) for $\sigma=\sqrt{{r_1}/{2}}$
    with $r_2 = -24\sigma^4=-6r_1^2$. Then one can find $\Lambda = \frac{3}{2}r_1 M_P^2$. This is a non-singular bouncing solution if $r_1>0$ with a de Sitter asymptotic for large $|t|$. It requires a radiation fluid.

   \item
   Bouncing Type II \cite{Koshelev:2012qn}:
   \begin{equation}\label{eq:superInf}
       a(t) = a_0 e^{\frac{\sigma}2t^2},~H=\sigma t,~R=126\sigma^2 t^2+6\sigma.
   \end{equation}
    This solution satisfies ansatz (\ref{eq:ansatz}) for $\sigma=-{r_1}/{6}$
    with $r_2 = 12\sigma^2={r_1^2}/{3}$. Then one can find $\Lambda = - {r_1}M_P^2/{12}$. This solution does not behave like de Sitter for all $t$ but leads to the bouncing behavior for $r_1<0$ thus requiring $\Lambda>0$ in the action. However, it requires no radiation.

    \item
    Bouncing type III:
    \begin{equation}\label{eq:bouSolsqrt}
        a(t) = a_0 \sqrt{\cosh{(\sigma t)}},~H=\frac\sigma2\tanh(\sigma t),~R=3\sigma^2.
    \end{equation}
    On this solution the background Ricci scalar is a constant throughout the evolution of the Universe. The Hubble parameter is however not. This solution does not require ansatz (\ref{eq:ansatz}) as such, {as $\Box R = 0$, and} the background behavior is very simple. One can find from (\ref{EOMtrace}) $\Lambda=3\sigma^2 M_P^2/4$. This is a non-singular bouncing solution if $\sigma$ is real, behaving like a de Sitter for large $|t|$. It requires a radiation fluid.

\end{enumerate}

To find out the radiation content  one can upon solving trace equation (\ref{EOMtrace}) substitute the found metric in the $(00)$ equation. For solutions obeying the ansatz the following neat expression can be obtained \cite{Koshelev:2016xqb}
\begin{equation}
    2\lambda\Fc_1\left[-\frac32 \dot H R+3H\dot R+9r_1H^2+\frac34r_2\right]=\rho_r=\rho_0\frac{a_0^4}{a^4},
\end{equation}
where $a_0$ {and $\rho_0$ are}  taken as the scale factor and radiation energy density at the time of bounce and also we take $t=0$ at the bounce. Bounce point is then characterized by $H=0$ and thus we come to a very short expression
\begin{equation}
    2\lambda\Fc_1\left.\left[-9 \dot H^2-\frac{3\Lambda r_1}{M_P^2}\right]\right|_{\text{Bounce}}=\rho_0.
    \label{interesting}
\end{equation}
Note that it is a general equation as long as a solution obeys ansatz (\ref{eq:ansatz}).

For the first bouncing solution (\ref{eq:bouSolcosh}) we get
\begin{equation}\label{eq:radDen}
     \rho_0 = -\frac{27}{2} \lambda r_1^2 \Fc_{1}.
\end{equation}
which would be a positive value if $\Fc_1<0$. Establishing whether it is possible is the question to be elaborate in Section~\ref{sec:ScalarPert}.
For the second bouncing solution (\ref{eq:superInf}) we get exactly $\rho_0=0$.
For the last bouncing solution  (\ref{eq:bouSolsqrt}) we use equations of motion (\ref{eq:eom}) directly as constants $r_{1,2}$  {remain arbitrary} even if one would try to impose an ansatz. Doing the algebra one yields
\begin{equation}\label{eq:radDen2}
    \rho_{0} = - \frac R4({M_P^2} + 2{\lambda} f_0 R),
\end{equation}
which would be a positive value if $({M_P^2} + 2{\lambda} f_0 R)<0$ (note that $R=3\sigma^2>0$). Again, establishing whether it is possible is the question to be elaborate in Section~\ref{sec:ScalarPert}.

The above are results derived in previous papers cited above. Our primary goal now is to connect those results and restrictions with a rigorous consideration of the DOF-s and corresponding new constraints. This is actually a step which has not been discussed in details before.

\section{Ghost-free form factors around Minkowski space-time}\label{sec:Mink}

DOF-s can be easily defined around Minkowski space-time or perhaps around a MSS background. Equations of motion (\ref{eq:eom}) clearly give Minkowski space-time as a solution if $\Lambda=0$ and (A)dS space-time if $\Lambda\neq0$. Minkowski space-time does not seem relevant for the main subject of this paper but it is  an instructive case study example.

(Perturbative) DOF-s are identified with the poles of the propagator. It should be only one pole to avoid the Ostrogradski ghost problem. Higher derivatives can make this happen only if they are forming an exponent of an entire function at some point.
Namely, as a starting point we say that a single DOF requires 2 initial conditions. For a general equation
\begin{equation}
 \prod_i^N(\Box-\mu_i^2)\phi=0,
\end{equation}
one needs $2N$ initial conditions and thus $\phi$ as a field essentially describes $N$ DOF-s. A subtle transition happens when the operator is an exponent of an entire function. Consider the following equation
\begin{equation}
 e^{\sigma(\Box)}\phi=0,
\end{equation}
If $\sigma(\Box)$ is an entire function then being in the exponent it results in a function which has no zeros on the whole complex plane. Such an operator can have no normalizable non-trivial solutions (see \cite{Vladimirov:1988bd} as well as a critical view on this question in \cite{Woodard:2026kuk}). Therefore, two equations
\begin{equation}
 \prod_i^N(\Box-\mu_i^2)\phi=0~\text{ and }~e^{\sigma(\Box)}\prod_i^N(\Box-\mu_i^2)\phi=0,
\end{equation}
yield an identical set of normalizable non-trivial solutions. A general solution of this class is given by
\begin{equation}
 \phi=\sum_i^N\phi_i~\text{ where }\phi_i\text{ satisfies }~(\Box-\mu_i^2)\phi_i=0,
\end{equation}
Thus any linear in $\Box$ factor results in a separate DOF.

Presence or absence of ghosts in AID gravity models was studied in \cite{Biswas:2016egy} in a covariant four-dimensional formalism using the York decomposition of the metric perturbation $g_{\mu\nu}\to g_{\mu\nu}+h_{\mu\nu}$  in different spins with respect to the four-dimensional symmetry group.
This approach does not account for constraints resolution in Einstein equations but instead must be accompanied with a spin projection for individual modes (see \cite{VanNieuwenhuizen:1973fi}). While the bottom line of that approach is positive and a ghost-free condition can be flawlessly achieved for a wide class of form factors, we are going to re-do this analysis in an $(1+3)$ formalism which more closely reflects the physics and singles out real propagating DOF-s in a more transparent way.

We proceed with determining the ghost-free conditions on the form factors $\Fc_R$ and $\Fc_C$ around Minkowski space-time. The perturbed Minkowski metric can be written as
\begin{equation}\label{eq:pertMink}
    ds^2 = a(\tau)^2\left[-(1+2\phi) d\tau^2 - 2 \partial_i \beta d\tau dx^i + \big\{(1 - 2 \psi) \delta_{ij} + 2 \partial_i \partial_j \gamma + 2\chi_{ij}\big\} dx^i dx^j\right].
\end{equation}
Note that on Minkowski space-time $a=1$ but we write it for generality to be used later for other backgrounds. Cosmic and conformal times are identical in Minkowski since $a=1$.
Here $\phi(\tau,\vec x), ~\psi(\tau,\vec x), ~\beta(\tau,\vec x), ~\gamma(\tau,\vec x)$ are scalar perturbations and $\chi_{ij}(\tau,\vec x)$ are tensor perturbations. Here we decompose perturbations into scalars, vectors and tensors with respect to the three-dimensional symmetry group which is always preserved for a FLRW metric (even if spatially curved). Vector perturbations are not particularly interesting as they break Lorentz invariance.
Tensor perturbations satisfies the transverse and traceless condition with respect to the three-dimensional space
\begin{equation}\label{eq:TT}
    \chi^i_i = \partial_i \chi^i_j = 0.
\end{equation}
In general perturbations are subject to gauge fixing which will be discussed below. However, the above defined tensor perturbations are gauge invariant.

Around Minkowski, equation of motion \eqref{eq:eom} can be re-written identically in the following form
\begin{equation}\label{eq:eomCommMin}
    \Big[ M_P^2 + 2\lambda \Box \Fc_C(\Box)\Big]G_{\nu}^{\mu} = 2 \lambda \Big(\Fc_R(\Box) + \frac{1}{3} \Fc_C(\Box)\Big) \big(\nabla^{\mu} \partial_{\nu} - \delta_{\nu}^{\mu} \Box\big) R + O(R^2).
\end{equation}
We do this by using relation (\ref{usefulDDWeyl}) with the aim of singling out explicitly all the terms linear in curvature, and to remove by product an explicit Weyl tensor in the linear in curvature terms as it is quite cumbersome to work with. Obviously only linear in curvature terms will give a non-trivial contribution to perturbations around Minkowski space-time.

Different spins do not mix in linear perturbations thus allowing to study them separately.
We skip many details in the main part of the paper for better readability. Useful technical equations used to derive the results presented below are accumulated in Appendix~\ref{app:perturbations}.

As usual tensor perturbations are technically simpler to tackle. Perturbing the above equation first only with tensor modes yields (where the fact that $\delta R=0$ on tensor perturbations greatly helps)
\begin{equation}\label{eq:tensorMink}
    \Big[M_P^2 + 2 \lambda   \Box \Fc_C (  \Box)\Big]   \Box \chi_j^i = \hat{M}_P^2   \Box \chi_j^i= 0.
\end{equation}
This equation of motion is non-trivially modified compared to the GR case.
A pure d'Alembertian factor reproduces the standard graviton in GR.
However due to the presence of the form factor $\Fc_C$ this equation can result in a spectrum plagued by ghost DOF-s on top of the massless spin-2 mode.
If the operator $\hat{M}_P^2$ can be made to be an exponent of an entire function such that $\hat{M}_P^2 = M_P^2 e^{2 \omega(\Box)}$, with $\omega(\Box)$ being an entire function, then no extra DOF-s arise even in the presence of the Weyl tensor. Then $\Fc_C$ can be written as
\begin{equation}\label{eq:FcMin}
    \Fc_C(\Box) = M_P^2 \frac{e^{2\omega(\Box)}-1}{2\lambda \Box},
\end{equation}
where analyticity of $\Fc_C(\Box)$ at zero requires $\omega(0) = 0$.
It is important to note that the above form of the form factor suggests that one can either have no  Weyl tensor dependent term at all, i.e. $\Fc_C=0$, or must arrange an infinite tower of derivatives in the form factor in order to avoid perturbative ghosts \cite{Stelle:1976gc,Ivanov:2016hcm,Clunan:2009er}.
As another note one sees that the consideration of tensor modes in no way touches the other form factor, $\Fc_R$, in Minkowski background.

At this step, we simply arranged a rescaling of the d'Alembertian in the linearized equation of motion (\ref{eq:tensorMink}) by an operator which is an exponent of an entire function. Hence no new states emerge.
The second order action variation is the most straightforward way to verify that existing excitations are not ghosts.
An explicit computation can be done by using the observation that tensor perturbations $\chi^i_j$ are a subset of covariant transverse and traceless perturbations ${h^\mu_\nu}^\perp$. This allows to utilize directly expression (\ref{d2properallRconstTENSORS}) for $\chi^i_j$. One readily gets
\begin{equation}
\delta^2_T S=\frac12\int
dx^4\sqrt{-  g}\ {\chi^i_j}\left[{M_P^2}+
2\lambda
  \Box\Fc_C(  \Box)\right] \Box{\chi^j_i}.
\label{d2properallRconst13t}
\end{equation}
The subscript $T$ indicates that this is the second order variation for tensor perturbations. Even though the operator in brackets does not give rise to new excitations we want it to be positive for real eigenvalues of the d'Alembertian to assure that it does not flip the sign of the linearized action on physically relevant values of momenta and thus a ghost like behavior does not emerge. To yield this property we must require that $\omega(\Box)$ is real valued for the real d'Alembertian eigenvalues.

To perturb equation (\ref{eq:eomCommMin}) with scalar modes we first introduce the gauge invariant combinations of the scalar perturbations $\Phi$ and $\Psi$ (see (\ref{GIvars})). In these variables perturbed $(00)$, $(0i)$ and $(i\neq j)$  components of \eqref{eq:eom} look respectively
\begin{eqnarray} \label{eq:00Min}
    \hat{M}_P^2 \Psi =  \hat{\lambda} \delta R, \\ \label{eq:0iMin}
    \hat{M}_P^2 \Psi\pr = \hat{\lambda} \delta R\pr, \\ \label{eq:ijMin}
    \hat{M}_P^2 (\Psi - \Phi) = 2\hat{\lambda} \delta R,
\end{eqnarray}
where $\delta R$ is presented in (\ref{deltaRscalars}),
operator $\hat M_P^2$ is exactly as in (\ref{eq:tensorMink}), and operator $\hat{\lambda}$ can be written as
\begin{equation}
    \hat{\lambda} = \lambda \Big[\Fc_R(\Box) + \frac{1}{3} \Fc_C(\Box)\Big].
\end{equation}
Here we use the spatial Fourier decomposition and thus $k$  is the co-moving wave-vector (see Appendix~\ref{appNotations} for details).

If $\Fc_R(\Box) + \frac{1}{3} \Fc_C(\Box) = 0$ then $\hat{\lambda} = 0$ even if $\lambda \neq 0$. Then equation \eqref{eq:ijMin} implies $\Phi = \Psi$, while \eqref{eq:00Min} implies $\Psi = 0$, i.e. the system does not have any propagating scalar DOF-s, which is similar to the Einstein's gravity.

The situation becomes more subtle if $\hat{\lambda} \neq 0$. Then equations \eqref{eq:00Min} and \eqref{eq:ijMin} give rise to the following condition
\begin{equation}
    \hat{M}_P^2 (\Phi + \Psi) = 0,
\end{equation}
and following the consideration of tensor modes and in particular equation \eqref{eq:FcMin}, which imposes that $\hat{M}_P^2$ should be an exponent of an entire function the last condition has only a trivial solution
$$\Phi + \Psi = 0.$$
With this constraint, equation \eqref{eq:00Min} can be re-written upon substituting $\delta R$ using (\ref{deltaRscalars}) as follows
\begin{equation}\label{eq:Wscalar}
    M_P^2 \Psi = 6 \lambda  \Box \Fc_{R}(  \Box) \Psi.
\end{equation}
and serves as a dynamical equation for scalar perturbations.

We see that scalar modes are driven by $\Fc_R(\Box)$ only and the above equation has the following important implications:
\begin{itemize}
    \item If $\Fc_R(\Box) = 0$ then $\Psi=0$, and no scalar DOF-s exist like in GR.
    \item If $M_P^2 -  6 \lambda   \Box \Fc_{R}(  \Box) = M_P^2 e^{2\sigma(  \Box)}$, with $\sigma(  \Box)$ being an entire function, then again $\Psi = 0$, and hence again no scalar DOF-s appear. For this to happen $\Fc_{R}$ should get the following form
    \begin{equation}\label{eq:noScalar2}
        \Fc_{R}(\Box) = -\frac{1}{3} M_P^2 \frac{e^{2\sigma(\Box)} - 1}{2 \lambda \Box}.
    \end{equation}
    Here, $\sigma(0) = 0$  such that $\Fc_{R}(\Box)$ is analytic at zero.
    \item If $M_P^2 -  6 \lambda   \Box \Fc_{R}(  \Box) = M_P^2 (1 -   \Box/\mu^2) e^{2\sigma(  \Box)}$ with $\sigma(\Box)$ being an entire function, then one propagating scalar DOF with mass $\mu^2$ appears. Indeed, for such a form factor equation \eqref{eq:Wscalar} effectively reduces to
    \begin{equation}\label{eq:psiMin}
        (  \Box - \mu^2) \Psi = 0.
    \end{equation}
    The corresponding form factor is
    \begin{equation}\label{eq:WscalarFF}
        \Fc_R(\Box) = -\frac{1}{3} M_P^2 \frac{ \Big(1 - \frac{\Box}{\mu^2} \Big) e^{2\sigma(\Box)} - 1}{2 \lambda \Box}.
    \end{equation}
    Analyticity of $\Fc_R(\Box)$ around zero requires $\sigma(0) = 0$.
\end{itemize}

In order to ensure that $\Psi$ is not a ghost DOF for the last choice of the form factor \eqref{eq:WscalarFF}, we need to derive the second order variation of the action and test the sign in front of the temporal part of the kinetic term for the field $\Psi$. A positive sign would mean that $\Psi$ is not a ghost. Using the constraint $\Phi = -\Psi$ along with equations \eqref{deltaRscalars} the second order action of $\Psi$ can be written as
\begin{equation}
    \delta_s^{(2)} {S} = -3 \int d \tau d^3x \Psi \big(M_P^2 - 6\lambda   \Box \Fc_R(  \Box)\big)   \Box \Psi.
    \label{d2Ss1try}
\end{equation}
The subscript $s$ indicates that this is the second order variation for scalar perturbations. This expression is obtained by a direct substitution of constraints and background quantities into (\ref{deltaEHADM}). Alternatively it could be ontained by utilizing expression (\ref{d2properallRconstSCALARSnobars}) which is a covariant second order action variation for a four-dimensional scalar. For the latter way one should compute $h$ as the trace of perturbations which in $(1+3)$ formalism is $h=2\Phi-6\Psi=-8\Psi$ upon resolving the constraints. This gives a factor $64$ which then exactly cancels $64$ in the dominator in (\ref{d2properallRconstSCALARSnobars}) and thus an exact matching with (\ref{d2Ss1try}) is restored.

We note here an important fact that in our scenario with higher (infinite) derivative operators the form factors themselves are not invertible. It is thus incorrect to use directly the $(00)$ equation as the Hamiltonian constraint to eliminate non-dynamical DOF-s. In other words, equation (\ref{eq:Wscalar}) should not be applied immediately to express $  \Box \Psi$ in the latter action like it would be done in standard cosmological perturbations in a local model. Instead, one can and should use equation (\ref{eq:psiMin}) which appears after the form factor is fixed. The latter equation simply gives $\Box\Psi=\mu^2\Psi$ and then using the required form of the form factor (\ref{eq:WscalarFF}) one can write
\begin{equation}
    \delta_s^{(2)} {S} =  -3 M_P^2 \int d \tau d^3 x \Psi \big(\mu^2 -   \Box\big)  e^{2\sigma(\Box)}\Psi.
    \label{ghostfreed2S}
\end{equation}
It is easy to see that in order to have a normal field, i.e. not a ghost, one must have $\sigma(\mu^2)$ to be real valued for real eigenvalues of the d'Alembertian. To reconfirm again that our logic of resolving constraints is correct we notice that our procedure is consistent the choice $\Fc_R=0$.

It is interesting that in this approach the mass of the scalar is not obligatory following from a coefficient in front of the local $R^2$ contribution. Indeed, expanding form factor (\ref{eq:WscalarFF}) in a Taylor series one gets
$$\Fc_R (\Box)=\frac{M_P^2}{6\lambda\mu^2}(1-2\sigma'(0)\mu^2)+O(\Box),$$
which would correspond to a standard $M_P^2/(12\mu^2)$ coefficient in front of $R^2$ in a local quadratic gravity which contains a scalar of mass $\mu$, only if the derivative of $\sigma(\Box)$ at zero vanishes.
Additionally, if one would try to organize a massless scalar excitation, that would result in a manifestly non-analytic form factor $\Fc_R$.

Our model can describe solutions which contain both Minkowski phase and some phase of evolution subject to full equations of motion with a non-trivial background. A notable example is the non-local inflation \cite{Koshelev:2020xby}. One can logically ask a question whether conditions (\ref{eq:ansatzImp}) dictated by ansatz are compatible with ghost-free conditions obtained above.
Therefore at this stage we want to examine all relevant constraints jointly. At present we are talking about a situation when Minkowski background is a solution, thus having $\Lambda=0$ and consequently $r_2=0$. Known bouncing solutions require a non-zero cosmological constant but it is illuminating to see how things work in Minkowski anyway. Following the steps already performed in \cite{Koshelev:2016xqb} we obtain as a result of full consideration that the following conditions must be satisfied
\begin{equation}
    r_1=\mu^2,~\text{ and }~\sigma(r_1)=0,
\end{equation}
This is a result of a direct evaluation of conditions (\ref{eq:ansatzImp}) on form factor (\ref{eq:WscalarFF}).
It implies that $r_1$ is essentially the mass squared of the scalar excitation and thus should be positive to avoid tachyons. Also one cannot have a situation with $r_1\neq0$ and no scalars around the Minkowski phase because a no scalars situation would correspond to $\mu^2=\infty$. We also emphasize once again that solutions with a pure Minkowski background do not require an ansatz and thus conditions (\ref{eq:ansatzImp}) should not be taken into account.

\section{Ghost-free form factors around de Sitter space-time}\label{sec:ScalarPert}

In this Section we extend an analysis performed above around the Minkowski background to the case of an de Sitter space-time. Namely, we want to find out perturbative physical (propagating) DOF-s around de Sitter space-time and combine important restrictions on form factors with conditions (\ref{eq:ansatzImp}) where relevant. While many stages of this analysis are applicable to the Anti-de Sitter space-time, our main attention is for cosmological solutions and thus we are primarily interested in either Minkowski or de Sitter asymptotics. We thus will turn to de Sitter only and will use the positivity of the Ricci scalar.

The perturbed FLRW metric can be written exactly as in equation (\ref{eq:pertMink}) with a difference that $a(\tau)$ is not a constant anymore. Rewriting equations of motion is not that helpful in this case though and one can just accomplish the task by directly evaluating perturbations starting from (\ref{eq:eom}) in their present form.

Like in Minkowski space-time we first start with the tensor modes. Equations for tensor modes $\chi_{ij}$ can be written as (where the fact that $\delta R=0$ on tensor perturbations greatly helps again)
\begin{equation}\label{eq:tensorDs}
    \Big[M_P^2 + 2 \lambda f_0   R + 2 \lambda \Big(  \Box -\frac{  R}{3}\big)\Fc_C\Big(  \Box+\frac{  R}{3}\Big)\Big] \Big(  \Box - \frac{  R}{6}\Big)\chi_j^i = \tilde{{M}}_P^2  \Big(  \Box - \frac{  R}{6}\Big)\chi_j^i = 0.
\end{equation}
In order to avoid any extra tensor DOF-s besides the usual massless spin-2 graviton operator $\tilde{{M}}_P^2$ should be an exponent of entire function $\tilde{{M}}_P^2 = M_P^2 e^{2\omega(\Box)}$ where $\omega(\Box)$ is an entire function, and as a result $\Fc_C (\Box)$ has the following form
\begin{equation}\label{eq:FcConstraintDs}
    \Fc_C (\Box) = M_P^2 \frac{e^{2 \omega (\Box)}-1-\frac{2\lambda f_0}{M_P^2}  R}{2 \lambda \left(\Box - \frac{2   R}{3} \right)}.
\end{equation}
The analyticity of $\Fc_C (\Box)$ at $\Box= \frac{2R}{3}$ requires, $2 \omega(\frac{R}{3}) = \log\big(1 + \frac{2 \lambda f_0}{M_P^2}R\big)$.
Proceeding to the second order action variation one gets
\begin{equation}
\delta^{(2)}_T S=\frac12\int
dx^4\sqrt{-  g}\ \chi^i_j\left[{M_P^2}+2\lambda {f_0}
  R+
2\lambda
\Fc_C\left(  \Box+\frac{  R}3\right)
\left( \Box-\frac{  R}3\right)\right]\left(   \Box-\frac {  R}6\right)\chi^j_i.
\label{d2properallRconst13tdS}
\end{equation}
The operator in brackets is required to be positive valued for real eigenvalues of the d'Alembertian, i.e. for real momenta, likewise in Minkowski case. To achieve this we require $\omega(\Box)$ to be real valued for real d'Alembertian eigenvalues. This additionally implies that
\begin{equation}
    1 + \frac{2 \lambda f_0}{M_P^2}  R>0.
    \label{positivityMink}
\end{equation}
Like in Minkowski case we note that the above shape of the form factor suggests that one can either have no the Weyl tensor dependent term at all, i.e. $\Fc_C=0$, or must arrange an infinite tower of derivatives in order to avoid perturbative ghosts \cite{Stelle:1976gc,Ivanov:2016hcm,Clunan:2009er}.
Also one sees that in the de Sitter background the tensor modes depend on the other form factor, $\Fc_R$, via its constant coefficient in the Taylor series expansion $f_0$ only.

In order to analyze scalar perturbations we as before utilize Bardeen potentials.
$(00)$, $(0i)$ and  $(i\neq j)$ components of perturbed equations (\ref{eq:eom})  can be written respectively as
\begin{eqnarray}\label{eq:00s}
    (M_P^2 + 2 \lambda f_0   R) \big(3\Hc^2 \Phi + 3\Hc \Psi \pr + k^2 \Psi\big) &=& \lambda k^2 \big(\Upsilon - 3 \Hc^2 \Upsilon + 3 \Hc \Upsilon\pr+\frac{2}{3} k^4 \Omega \big), \\ \label{eq:0is}
     (M_P^2 + 2 \lambda f_0   R) (\Psi\pr + \Hc \Phi) &=& 2\lambda\big(\Upsilon\pr -\Hc \Upsilon + \frac{2}{3} k^2 \Theta \big) ,\\ \label{eq:ijs}
    (M_P^2 + 2 \lambda f_0   R) (\Psi - \Phi) &=& 2\lambda \Upsilon - 2\lambda \big( \Theta\pr + 2 \Hc \Theta + \frac{k^2}{3} \Omega \big),
\end{eqnarray}
where
\begin{eqnarray}
    \Theta = \Omega\pr + \Hc \Omega, ~ \Omega  = \Fc_C(\Box + 6 H^2) \frac{\Phi + \Psi}{a^2},~\Upsilon = \Fc_R(\Box) \delta R.
\end{eqnarray}
Noting that on de Sitter $\Hc^2 = \Hc\pr$ and $R = 12\Hc^2/a^2$, the above system of equations leads to the following condition
\begin{equation}\label{eq:scalarConstraint}
   \Big[(M_P^2 + 2 \lambda f_0   R) + 2 \lambda \Big(  \Box - \frac{  R}{6}\Big) \Fc_C \Big(  \Box +\frac{  R}{2} \Big)\Big] \Big(\frac{\Phi+\Psi}{a^2}\Big) = \tilde{{M}}_P^2\vert_{  \Box \rightarrow   \Box +\frac{  R}{6}} \Big(\frac{\Phi+\Psi}{a^2}\Big) = 0.
\end{equation}
Here the operator acting on the combination $(\Phi+\Psi)$ is exactly the same operator as the one acting on the tensor modes in equation \eqref{eq:tensorDs} up to a shift $  \Box \rightarrow   \Box +\frac{  R}{6}$. Since $\tilde M_P^2$ was restricted to be an exponent of an entire function, the latter equation has only a trivial solution of
$$\Phi+\Psi = 0.$$
Upon using these condition, the $(i\neq j)$ equation \eqref{eq:ijs} can be re-written as
\begin{equation}\label{eq:scalarDs}
    (M_P^2 +2 \lambda f_0   R) \Psi = 2 \lambda (  R + 3   \Box) \Fc_R(  \Box)\Psi.
\end{equation}

We see that likewise in Minkowski case scalar modes are driven by $\Fc_R(\Box)$ only and the above equation has the following important implications:
\begin{itemize}
    \item If $\Fc_R(\Box) = 0$ then $\Psi=0$, and no scalar DOF-s exist like in GR with a cosmological term.
    \item
    $(M_P^2 +2 \lambda f_0   R) - 2\lambda (  R + 3   \Box) \Fc_R(  \Box) = M_P^2 e^{2\sigma(\Box)}$, with $\sigma(\Box)$ being an entire function: this implies that \eqref{eq:scalarDs} has only the trivial solution $\Psi = 0$ and thus no propagating scalar DOF-s. The corresponding form factor is
    \begin{equation}\label{eq:noScalarDs}
         \Fc_R(\Box) = -\frac{M_P^2}{3} \frac{e^{2 \sigma(\Box)}-1 - \frac{2 \lambda f_0 }{M_P^2}R}{2 \lambda \left(\Box + \frac{R}{3}\right)}.
    \end{equation}
    Here the analyticity of $\Fc_R(\Box)$ at $  \Box =-{  R}/{3}$ requires $2\sigma(-{  R}/{3}) = \log(1+ \frac{2 \lambda f_0}{M_P^2}  R)$. Note that condition (\ref{positivityMink}) implies that $\sigma(-  R/3)$ is real. In general this is analogous to the situation without scalars around Minkowski background for form factor (\ref{eq:noScalar2}). Moreover, the form factor normalization requires that $\sigma(0)=0$, and on contrary to the Minkowski case this is not dictated by the form factor analyticity.
    \item
    $(M_P^2 +2 \lambda f_0   R) - 2\lambda (  R + 3   \Box) \Fc_R(  \Box) = M_P^2 (1- {  \Box}/{\mu^2} )e^{2\sigma(  \Box)}$, with $\sigma(  \Box)$ being an entire function. This results in one propagating scalar degree of freedom with mass $\mu^2$. This allows to replace equation \eqref{eq:scalarDs} with
    \begin{equation}\label{eq:massiveScalar}
        (  \Box - \mu^2) \Psi = 0,
    \end{equation}
    and the corresponding form factor is
    \begin{equation}\label{eq:ff1}
    \Fc_R(\Box) = -\frac{M_P^2}{3} \frac{\left(1 - \frac{\Box}{\mu^2}\right) e^{2 \sigma(\Box)}-1 - \frac{2 \lambda f_0 }{M_P^2}R}{2 \lambda \left(\Box + \frac{R}{3}\right)}.
    \end{equation}
    The analyticity of $\Fc_R$ at $  \Box = -  R/3$ requires
    $2\sigma(-  R/3)=\log\left(\frac{3\mu^2}{M_P^2}\frac{M_P^2+{2\lambda f_0}  R}{3\mu^2+   R}\right)$.
    This situation is analogous to the Minkowski space-time accompanied with one scalar degree of freedom with form factor \eqref{eq:WscalarFF}. Again, the form factor normalization implies $\sigma(0)=0$.
    Moreover, like in Minkowski case, if one would try to organize a massless scalar excitation keeping the tensor modes healthy (i.e. satisfying condition (\ref{positivityMink})), that would result in a manifestly non-analytic form factor $\Fc_R$.
\end{itemize}

It is important to test the last choice of the form factor (\ref{eq:ff1}) against the second variation of the action in order to see whether the resulting DOF is a ghost or not.
Repeating steps like in the analysis around Minkowski space-time in the previous Section and especially using the simplifying relation $\Phi=-\Psi$ in order to deduce the second order variation of the action for a single degree of freedom one can proceed by utilizing expression (\ref{d2properallRconstSCALARSnobars}) with $h=-8\Psi$. The second order action variation the will take the form
\begin{equation}\label{eq:scalarDsAction}
\delta^{(2)}_s S=-3M_P^2\int
dx^4\sqrt{-  g}\ \Psi\left(  \Box+\frac{  R}3\right)\frac1{\mu^2}(\mu^2-  \Box)e^{2\sigma(  \Box)}\Psi.
\end{equation}
where we used the explicit expression for form factor (\ref{eq:WscalarFF}). Now, resolving the Hamiltonian constraint via equation (\ref{eq:massiveScalar}) one comes to the following action
\begin{equation}
\delta^{(2)}_s S=-3M_P^2\left(1+\frac{  R}{3\mu^2}\right)\int
dx^4\sqrt{-  g}\ \Psi(\mu^2-  \Box)e^{2\sigma(  \Box)}\Psi.
\label{d2Sscalars1mode}
\end{equation}
This gives a normal, i.e. not a ghost degree of freedom if $\left(1+\frac{  R}{3\mu^2}\right)>0$ and $\sigma(\Box)$ is real valued for real eigenvalues of the d'Alembertian, or conversely $\left(1+\frac{  R}{3\mu^2}\right)<0$ and $\sigma(\Box)=r\pm\pi i/2$ where $r$ is real. The former case corresponds to a normal field or to a heavy tachyon with $\mu^2<-   R/3$.  The latter case corresponds to a light tachyon with $-  R/3<\mu^2<0$. Since tachyons lead to instabilities, their presence is not immediately interesting overall.

\section{Ghost-free bounces?}
\label{ghostfreeyesno}

We have exact bouncing solutions in our model which have a de Sitter asymptotic. One can ask whether conditions (\ref{eq:ansatzImp}) dictated by ansatz are compatible with ghost-free conditions in de Sitter space-time obtained in the previous Section.
To have a de Sitter asymptotic we must have $\Lambda\neq0$ and consequently $r_2\neq0$ as well.
Moreover we consider solutions which are not a pure de Sitter, or not as simple as $R=\const$ when a simplifying ansatz does not lead to non-trivial consequences. Thus, for instance $\Box R=0$ situation happens in at most countable set of non-accumulating points. This assures that only in the asymptotic or in some discrete set of time values $R\to -r_2/r_1$.
As a result of joint consideration of all conditions and constraints, we obtain that the following extra relations must be satisfied:
\begin{itemize}
    \item
    Considering the second condition in (\ref{eq:ansatzImp}) and canceling obviously identical terms (under an assumption that $r_1\neq0$) and using that in the de Sitter asymptotic $R\to -r_2/r_1$ we arrive at an expression
    \begin{equation*}
    \frac R3\frac{1-\frac{r_1}{\mu^2}}{r_1+\frac R3}=-r_1\frac{1-\frac{r_1}{\mu^2}}{r_1+\frac R3}.
    \end{equation*}
    Assuming that $R$ can be equal $-r_1/3$ only in a countable (perhaps an empty) set of time points, we can cancel $r_1+R/3$ factor as well and thus come to the conclusion that
    $$r_1=\mu^2.$$
    This resembles the Minkowski case and also underlines that one cannot come up with a solution which satisfies ansatz for a finite $r_1$ and has no a scalar mode as $r_1$ is essentially the mass of that scalar.
    \item
    Evaluating $\Fc(r_1)$ we get taking into account that $r_1=\mu^2$
    \begin{equation}
        \Fc_1=\frac{M_P^2 + {2 \lambda f_0 }R}{2 \lambda \left(3\mu^2 + {R}\right)}.
        \label{scalarghostindspre}
    \end{equation}
    According to the discussion after (\ref{d2Sscalars1mode}) for a non-tachyonic scalar mode with $\mu^2>0$ (and also for a light tachyon with $-R/3<\mu^2$) we get
    \begin{equation}
    \Fc_1>0\label{scalarghostinds},
    \end{equation}
    where the inequality is strict. For a heavy tachyon with $\mu^2<-R/3$ we will get
    $
    \Fc_1<0\label{scalarghostindstach}
    $.
    \item
    The first condition in (\ref{eq:ansatzImp}) results in an extra constraint
    \begin{equation}\sigma\left(-\frac R3\right)=\sigma(r_1).\end{equation}
\end{itemize}

Now we are going to examine what does the analysis before this point mean for bouncing solutions presented in Section~\ref{sec:AIDintro}.
\begin{itemize}
\item
Solution (\ref{eq:bouSolcosh}) requires radiation. Unfortunately, this radiation will be a ghost  according to formula (\ref{eq:radDen}) due to the fact that for this solution $r_1=\mu^2>0$ and therefore $\Fc_1>0$.
\item
Solution (\ref{eq:bouSolsqrt}) does not require an ansatz as it features $R=\const$. However, according to formula (\ref{eq:radDen2}) a required radiation will be a ghost as long as condition (\ref{positivityMink}) is applied. The latter is required for the analyticity of the $\Fc_C(\Box)$ form factor at point $\Box=-R/3$.
Non-analyticity of the form-factor at least in one finite point on the complex plane would mean that the form factor is not an entire function anymore and one cannot easily work with them using the series representation everywhere even on a positive real axis as the radius of convergence of the Taylor series becomes finite. It leaves a not obvious window for a non-singular and a ghost-free bounce but extending our analysis to non-entire form factors is beyond the scope of the present paper.
\item
Solution (\ref{eq:superInf}) does represent a bounce but it does not feature a de Sitter asymptotic. Even though it requires a cosmological constant, it is not clear why anything should be imposed on this solution following conditions obtained around a de Sitter background which is not related to the bouncing solution at all. This can open another window to a possible ghost-free configuration but the super-inflation behavior of this solutions makes it not clearly suitable for cosmological scenarios.
\end{itemize}

We then can come to an even more general conclusion regarding the bounce and ghosts if a solution satisfies the ansatz (\ref{eq:ansatz}). Returning to formula (\ref{interesting}) which computes the energy of radiation at the bounce point we can rewrite it as
\begin{equation}
    2\lambda\Fc_1  \left.\left[-9 \dot H^2-\frac{3\Lambda \mu^2}{M_P^2}\right]\right|_{\mathrm{Bounce}}=\rho_0.
    \label{interestingmore}
\end{equation}
where we used relation (\ref{scalarghostindspre}) and the established above fact that $r_1=\mu^2$. For a non-tachyonic scalar mode with $\mu^2>0$ we also got $\Fc_1>0$ and the only chance of getting non-ghost radiation is to have a negative cosmological constant. A situation not clearly compatible with the de facto observed Universe. A light tachyon with $-R/3<\mu^2$ results in $\Fc_1>0$ and will make the radiation not a ghost but the tachyon is unstable on its own. Another possibility is to have a heavy tachyon with $\mu^2<-R/3$ which gives $\Fc_1<0$ but then $\dot H$ should be large at the bounce point to make an expression inside of the brackets negative.

To conclude this Section, we clearly see that ghosts for bouncing solutions in this model can be avoided either via an introduction of a tachyonic degree of freedom or via a not easily justifiable negative cosmological constant. Solution (\ref{eq:superInf}) is not cosmologically viable as some mechanism is needed to stop the super-inflation and this requires additional ingredients beyond this model itself.

\section{Analysis of the scalar and tensor modes of perturbations}
\label{sec:spectra}
In this section we will be discussing the dynamics of scalar and tensor perturbations around the bouncing solutions of \eqref{eq:bouSolcosh} and \eqref{eq:bouSolsqrt}. Both of these solutions have a de Sitter asymptotic. However for both of these solutions presence of a ghost radiation is inevitable following the analysis in previous Sections. Having in mind the arguments from the Introduction that the presence of ghosts does not eliminate the relevance of studying the bounce we proceed with looking for observational implications of our approach.

\subsection{The scalar modes}
To analyze the scalar modes we first reiterate that even though a radiation has to be present for both bouncing solutions under consideration, around de Sitter limit of both the solutions the radiation energy density is negligible compared to the total energy density of the Universe. As a consequence radiation only generates iso-curvature perturbations and is completely decoupled from the curvature perturbation generated by the gravity sector \cite{Gordon:2000hv}.

We expect that due to the fact that the scalar mode is massive, we only need to analyze its behavior at late times with an expectation that it  will decay anyway. As shown in the previous section around de Sitter space-time the dynamics of scalar DOF is governed by the action \eqref{d2Sscalars1mode}. We introduce the canonically normalized variable $v (x,\tau)= \sqrt{6(1+R/3\mu^2)} \cdot M_Pae^{\sigma(\Box)} \Psi (x,\tau)$ and its equation of motion in terms of Fourier transformed variable $v_k (\tau) = v(\tau,\Vec{k})$ can be written as
\begin{equation}\label{eq:massScalar}
    v_k\prt (\tau) + k^2 v_k(\tau) + \Big\{\mu^2a^2 - \frac{a\prt}{a}\Big\} v_k(\tau)=0.
\end{equation}
Around the de Sitter limit the scale factor is $a(\tau) = - \frac{1}{H \tau} = -\frac{1}{H_{dS}}$, where $H_{dS} = H(\vert t \vert \rightarrow \infty)$ and the solution to the above equation  can be written as
\begin{equation}\label{eq:scalarMode}
    v_k(\tau) = \sqrt{-\tau} \Big[c_1 H_{\nu}^{(1)}(-k \tau) + c_2 H_{\nu}^{(2)}(- k \tau)\Big],
\end{equation}
$H_{\nu}^{(1),(2)}$ are the Hankel functions of first and second kind, and $\nu = \sqrt{\frac{9}{4} - \frac{\mu^2}{H_{dS}^2} }$, while $c_1$ and $c_2$ are constants. Upon choosing the Bunch-Davies vacuum the constants can be determined as $c_2 = 0$ and $c_1 = \frac{\sqrt{\pi}}{2}$ (ignoring some unimportant phase factors). With this the  solution for the Fourier transform of $\Psi(x,\tau)$ can be written as
\begin{equation}\label{eq:scalarSol}
    \Psi_k(\tau) = \sqrt{\frac{\pi}{24}} \frac{1}{M_P\Big(1 + \frac{R}{3 \mu^2}\Big)^{1/2}} \frac{e^{-\sigma(\mu^2)}}{a(\tau)} \sqrt{\tau} H_{\nu}^{(1)} (-k\tau),
\end{equation}
here we evaluate $\Psi_k(\tau)$ (equivalently $\Psi(\tau,\Vec{k})$) for modes that satisfy \eqref{eq:massiveScalar} such that the argument of $\sigma$ is $\mu^2$.  In the previous Section we have already shown that the mass of $\Psi(x,\tau)$ is  $\mu^2 = r_1$, and this has a non-trivial consequence for the observability of scalar modes in CMB. For the solution \eqref{eq:bouSolcosh} we can determine $r_1 = 2 H_{dS}^2$, and hence $\nu = 1/4$. Now the asymptotic behavior of Hankel function suggests that $H^{(1)}_{1/4}(-k \tau \ll 1) \sim (-k\tau)^{-1/4}$ and hence a super-horizon mode of $\Psi_k $ would scale as $ (-k \tau)^{5/4}$ implying that at late times the scalar mode $\Psi_k$ will decay. For solution \eqref{eq:bouSolsqrt} we note that $r_1$ and hence the mass of the mode $\mu$ is undetermined. If $\mu^2 \ll H_{dS}^2$, $\Psi_k$ can be scale invariant but its amplitude will have a suppression factor of $\sqrt{(3\mu^2/R)}$ which is small. If $\mu^2\geq H_{dS}^2$ the mode will decay at late times anyway. We also note that one can avoid a scalar mode for solution (\ref{eq:bouSolsqrt}) altogether since $\mu^2$ is unconstrained and in principle can be formally taken infinite.

\subsection{The tensor modes}
The second variation of the action \eqref{eq:masterL} for the tensor modes around MSS background can be written according to equation \eqref{d2properallRconst13tdS}.
 Here we can see that the dynamics of $\chi_{ij}$ is governed by the operator $M_P^2 e^{2\omega(\Box)} (\Box-R/6)$, and as $e^{2\omega(\Box)}$ is an invertible operator we further introduce a canonically normalized variable $\tilde{\chi}_{ij}(x,\tau) = \sqrt{2}a e^{\omega(\Box)} \chi_{ij}(x,\tau)$. This variable can further be expanded in Fourier modes denoted as $\tilde{\chi}_k(\tau)= \tilde{\chi}(\tau,\Vec{k})$ and its equation of motion can be written as
\begin{equation}\label{eq:MST}
    \tilde{\chi}_k^{\prime \prime}(\tau) + \Big(k^2 - \frac{a\prt}{a}\Big)\tilde{\chi}_k(\tau) = 0.
\end{equation}
Here we note that the tensor modes  have two polarizations, and thus computing the spectrum we need to take a sum over these two polarizations modes.
Since on contrary to scalar modes in the previous Subsection, the graviton is massless and is relevant throughout the whole evolution, we need to solve this equation in all three phases of the Universe evolution. Phase~1 is the de Sitter like contraction phase; Phase~2 is the bouncing phase where $H\approx0,~\dot{H}>0$; Phase~3 is the de Sitter like expansion phase which is symmetric to the contraction phase. Again dynamics of the tensor modes will be non-trivial around bouncing phase such that the form factors would introduce DOF-s with complex masses squared \cite{Tokareva:2024sct}. For the estimation of the tensor power-spectrum in the current bouncing scenarios, we assume that  if these extra modes appear, they will be classically decaying and in this article we will neglect their effects. We recall that such a behavior is subject to the form factor shapes which determine the model.

The bouncing scenario of \eqref{eq:bouSolsqrt} is simpler to analyze since it features $R=\const$. Also, it is similar to the case of \cite{Durrer:2002jn,Cai:2008qw} and we can apply similar matching conditions on the boundary of the three phases described above. In our case we will be matching the solutions of \eqref{eq:MST} and its derivative on $-t_B$ and $t_B$ surfaces where the time $\vert t_B \vert$ denotes the time when $\dot{H}$ is large enough satisfying $\dot{H}/H^2 = 1$.
 Considering $H_{dS} = 10^{-5} M_{P}$, the time $\vert t_B\vert$ can be computed as
$\vert t_B \vert = 1.15/(2  H_{dS})\simeq 5.7 \times 10^4 M_P^{-1}$.
We introduce conformal times $\tau_{B-}$ and $\tau_{B+}$  corresponding to $\mp t_B$, and conformal Hubble parameters $\Hc_{B-}$ and $\Hc_{B+}$ corresponding to Hubble parameters at $\mp t_B$ following \cite{Cai:2008qw} . As the bounce is symmetric we have $\Hc_{B+} = -\Hc_{B-}$.

 In the contraction phase \eqref{eq:MST} has a solution
\begin{equation}
    \tilde{\chi}_k^C(\tau) = \sqrt{\tau - \tilde{\tau}_{B-}}\Big\{A_k^C H_{3/2}^{(1)}[-k (\tau-\tilde{\tau}_{B-})]+B_k^C H_{3/2}^{(2)}[-k (\tau-\tilde{\tau}_{B-})]\Big\},
\end{equation}
where $\tilde{\tau}_{B^-} = \tau_{B-} - \frac{2}{\Hc_{B-}}$, and the superscript $C$ denotes contraction phase. Using the Bunch-Davies initial condition one can compute the constants as $A_k^C = \sqrt{\pi}/2 e^{-i \frac{\pi}{2}}, ~B_k^C = 0$ and the solution for modes outside the Hubble radius can be written as
\begin{equation}
    \tilde{\chi}_k^C(\tau) = \frac{-i}{\sqrt{2}}\frac{1}{k^{3/2}} (\tau - \tilde{\tau}_{B-})^{-1}.
\end{equation}
In the time interval from $\tau_{B-}$ to $\tau_{B+}$ one can check that $a\prt/a \approx 2 H_{ds}^2$,
hence the solution for modes with $k< H_{ds}$ is
\begin{equation}
    \tilde{\chi}_k^B (\tau) = C_k^B e^{\big\vert \sqrt{k^2 - 2H_{ds}^2}\big\vert (\tau - \tau_B) } + D_k^B  e^{-\big\vert \sqrt{k^2 - 2H_{ds}^2}\big\vert (\tau - \tau_B)}.
\end{equation}
Here $\tau_B$ corresponds to the conformal time at bounce, and superscript $B$ is used to denote bouncing phase. At last the solution in the expanding phase can be written as
\begin{equation}
    \tilde{\chi}^{E}_k (\tau) = \sqrt{\tau - \tau_{B+}} \Big\{E_k^E J_{-3/2}[k(\tau - \tilde{\tau}_{B+})]+F_k^E J_{3/2}[k(\tau - \tilde{\tau}_{B+})]\Big\} \simeq \sqrt{\frac{2}{\pi}}\frac{F^E_k}{3}k^{3/2} (\tau - \tilde{\tau}_{B+})^2,
\end{equation}
where $\tilde{\tau}_{B+} = \tau_{B+} - \frac{2}{\Hc_{B+}}$, the superscript $E$ denotes the expansion phase, and $J_{\nu}(x)$ is the Bessel function of first kind. In the right most expression we used an asymptotic expansion of the Bessel function such that the term proportional to $E^E_k$ becomes negligible and only the term corresponding to $F^E_k$ survives.

Upon matching the mode functions and its first derivative at the $\tau_{B-}$ and $\tau_{B+}$ surfaces we can determine $F^E_k$ as
\begin{eqnarray}
    F^E_k = - \frac{3 i \sqrt{\pi}}{32}\frac{\Hc_{B+}^3}{k^3} \Bigg\{-2 \cos\varphi
    + \frac{\Hc_{B+}}{\vert\sqrt{k^2 - 2H_{ds}^2}\vert} \sin\varphi\Bigg\}\text{ with }\varphi=\Big\vert\sqrt{k^2 - 2H_{ds}^2}\Big\vert (\tau_{B-} - \tau_{B+}).
\end{eqnarray}
We can check that for $H_{ds} = 10^{-5}$ and for modes $k \ll H_{ds}$, the quantity $\vert\sqrt{k^2 - 2H_{ds}^2}\vert (\tau_{B-} - \tau_{B+}) \sim 1.61$ on solution \eqref{eq:bouSolsqrt}. Here we also used $H_{B+} \approx H_{ds}$ which is evident from the behavior of \eqref{eq:bouSolsqrt}.  With these considerations one can reduce
$
    F^E_k \approx -0.14 i {\Hc_{B+}^3}/{k^3},
$
and hence
\begin{equation}
    \tilde{\chi}^E_k (\tau) = -\sqrt{\frac{2}{\pi}}\frac{0.14 i}{3}\frac{\Hc_{B+}^3}{k^{3/2}}(\tau - \tau_{B+})^2,
\end{equation}
which leads to the following power-spectrum of the tensor mode functions (for $\lambda =1$) at late times
\begin{equation}
    P_T = 2 \frac{k^3}{2 \pi^2} \vert \chi_k \vert^2 \approx 0.14\cdot10^{-3} \frac{H_{dS}^2}{M_P^2} e^{-2\omega(R/6)}.
\end{equation}
 Here $\chi_k$ is the Fourier transform of $\chi_{ij}(x,\tau)$ and the factor of 2 in the definition of $P_T$ appears due to the sum over two polarizations. Here the modes satisfy \eqref{eq:tensorDs} such that the argument of $\omega$ is $R/6$.  From the analyticity requirement of $\Fc_C$ at $\Box =2R/3$ the entire function $\omega$ was constrained at $\Box = R/3$. Now we can see that $\omega$ also has to be constrained at $\Box = R/6$ from the observable bounds on tensor modes.

For the solution \eqref{eq:bouSolcosh} a similar analysis can be performed to determine the spectrum. Dor this solution $H_{B+} = 10^{-5} H_{dS}$ for $H_{dS} = 10^{-5} M_P$. Using the previous algorithm one can compute $F^E_k = \frac{0.33 i}{k^3} \Hc_{B+}^3$, and again the spectrum is proportional to $e^{-2\omega(R/6)}$, signaling additional constraint on $\omega$ has to be used in order to satisfy the observations.

\section{Summary}\label{sec:conclusion}
AID gravity theories are a ghost-free and renormalizable description of gravity around at least covariantly constant space-times as long as the perturbative spectrum replicates the one of Einstein's GR, that is, only a massless 2-polarization graviton is present. Extra DOF-s, like scalar modes or fluids can in principle change this neat picture. Within these theories non-singular bouncing cosmology can be realized as exact solutions to equations of motion. Known solutions (see (\ref{eq:bouSolcosh}) and (\ref{eq:bouSolsqrt})) require a positive cosmological constant and perhaps radiation, and they satisfy an ansatz (\ref{eq:ansatz}) (or are very simple having, say, $R=\const$). We have found out that in general one cannot avoid a presence of some unstable mode, a ghost or a tachyon, but not both simultaneously, for a non-negative cosmological constant. However, the presence of a negative cosmological constant seems in contradiction with existing observations. We note that known solutions within our model either have a ghost radiation or do not have a de Sitter asymptotic at all (see (\ref{eq:superInf})) making their interpretation even more difficult.

In this paper, we have performed a detailed analysis of DOF-s around Minkowski space-time as a warm up exercise, and then around de Sitter space-time, both in $(1+3)$ formalism. We did this analysis for both tensor and scalar modes (vector modes vanish on MSS backgrounds). This is a study which was not performed before in full generality for infinite derivative theories. We then explore how the ghost-free conditions around those MSS backgrounds can be joined with the conditions (\ref{eq:ansatzImp}) dictated by the used ansatz. This step reveals that solutions which are subject to ansatz must have a scalar degree of freedom. We then looked more closely into known bouncing solutions.

We have found for solution \eqref{eq:bouSolcosh} which is subject to the ansatz that the required radiation must be a ghost. For solution (\ref{eq:bouSolsqrt}) which has $R=\const$ and thus does not need an ansatz we have found that an attempt to make the required radiation not a ghost results in the loss of the property of the form factor $\Fc_C(\Box)$ to be an entire function. The latter situation means that one cannot strictly speaking use a Taylor expansion for form factors.
Solution (\ref{eq:superInf}) is also a bouncing solution but it describes a  super-inflating Universe and does not seem to be an interesting scenario. On this way we come to a conclusion discussed around formula (\ref{interestingmore}) that a healthy regular bounce without ghosts or tachyons can be realized only with a negative cosmological term in the action. We thus emphasize these two narrow possibilities to avoid instabilities in bouncing scenarios still keeping the cosmological term positive: non-analytic form factors, or a bounce which leads to a super-inflation. These cases are not covered in the current analysis and present interesting questions for future publications.

Leaving the resolution of instabilities usual for many bounce models for future research, we have computed the scalar and tensor perturbations around the bouncing solutions \eqref{eq:bouSolcosh} and \eqref{eq:bouSolsqrt} which behave as de Sitter space-time at late times. However, these solutions do not provide a graceful exit from the de Sitter expansion. The computed scalar and tensor modes presented in Section~\ref{sec:spectra} thus can not explain the CMB observations. To address the latter, we need instead to consider a bouncing  solution where a graceful exit is present. Nevertheless with our computations we capture the general characteristics and  leading order effects of the perturbations which will also be present in an observationally relevant bouncing solution but with corrections. With these considerations
we arrive at the following conclusions:
\begin{itemize}
    \item Our analysis of the form factors around de Sitter space-time reveals that the theory consist of a massive scalar degree of freedom.
        The behavior of the scalar perturbation is derived by only considering its characteristics around the de Sitter limit, and its full evolution across the contraction, bounce and expansion phases of the Universe would be much more involved. But as discussed in the course of the paper the radiation dilutes quickly once the Universe enters de Sitter expansion phase, and extra corrections which may appear around the bouncing phase would vanish once the de Sitter limit is reached.
    As a result if a scalar mode is heavy it would decay at late times, if it is light, its amplitude would be very small. In both cases such a scalar can not produce observable temperature fluctuations of CMB. So to explain the observed temperature fluctuations curvaton like \cite{Enqvist:2001zp,Bozza:2002fp,Cai:2011zx} or a similar mechanism is needed where iso-curvature perturbations of light fields get converted to curvature perturbations at late times.
    \item
    Upon computing the tensor power-spectrum we find that around these solutions the amplitude has a factor of $e^{\omega(R/6)}$ which can enhance or suppress the signal depending on the behavior of $\Fc_C(\Box)$ at ${R}/{6}$. This can be considered as a constraint on $\Fc_C(\Box)$ from observations.

\end{itemize}

To summarize, in this paper our primary focus was to explore DOF-s appearing in de Sitter asymptotics of known bouncing solutions of AID gravity. It turns out that instabilities occur inevitably in bouncing scenarios despite the success of this approach of constructing a renormalizable gravity theory around covariantly constant backgrounds. Moreover, existing scenarios are physically incomplete as they lack of a mechanism that allows the Universe to smoothly transit to radiation or matter domination phase after bounce. This is required to explain the observed CMB scalar fluctuations. We aim to address these issues in forthcoming papers.

\section*{Acknowledgments}
AK and AN are extremely grateful to very inspiring discussions with Robert Brandenberger on the subject of bounce, and to Masahide Yamaguchi for various discussions about higher derivative cosmological models.

\appendix

\section{Notations and useful formulae}
\label{appNotations}

The dimension of the space-time is 4 and the signature is $(-+++)$.
The spatially flat FLRW Universe metric is
\begin{eqnarray}
ds^{2}=-dt^{2}+a(t)^{2}\left({dr^{2}}+r^{2}d\Omega^{2}\right)\,.\label{frwcosmic}
\end{eqnarray}
$t$ is the cosmic time and $a(t)$ is the scale factor. We work with zero spatial curvature only.
Our conventions are
\begin{eqnarray}
\Gamma_{\mu\nu}^{\rho}=\frac{1}{2}g^{\rho\sigma}(\pd_{\mu}g_{\nu\sigma}+\pd_{\nu}g_{\mu\sigma}-\pd_{\sigma}g_{\mu\nu}).
\end{eqnarray}
\begin{equation}
R_{\mu\nu\rho}^{\sigma} = \pd_{\nu}\Gamma_{\mu\rho}^{\sigma}-\pd_{\rho}\Gamma_{\mu\nu}^{\sigma}+\Gamma_{\chi\nu}^{\sigma}\Gamma_{\mu\rho}^{\chi}-\Gamma_{\chi\rho}^{\sigma}\Gamma_{\mu\nu}^{\chi},\quad
R_{\mu\nu}=R_{\mu\sigma\nu}^{\sigma},\quad R=R_{\mu}^{\mu}\label{def:ricci}
\end{equation}
The covariant derivative and the d'Alembertian are defined as
\begin{eqnarray}
\cpd_{\mu}F_{.\beta.}^{.\alpha.}=\pd_{\mu}F_{.\beta.}^{.\alpha.}+\Gamma_{\mu\chi}^{\alpha}F_{.\beta.}^{.\chi.}-\Gamma_{\mu\beta}^{\chi}F_{.\chi.}^{.\alpha.}\,,\quad\Box=g^{\mu\nu}\cpd_{\mu}\cpd_{\nu}\,.
\end{eqnarray}

For metric (\ref{frwcosmic}) we have
\begin{eqnarray}
  \Gamma_{0i}^{j} & = & H\delta_{i}^{j},\quad  \Gamma_{ij}^{0}=g_{ij}H,\quad H=\dot a/a,\\
  R & = & 12H^2+6\dot H,\quad   R_{\mu\nu}=\left(\begin{array}{cc}
-3(H^2+\dot H) & 0\\
0 & (3H^2+\dot H)g_{ij}
\end{array}\right),\\
  G_{\mu\nu}&=&\left(\begin{array}{cc}
3H^2 & 0\\
0 & -(3H^2+2\dot H)g_{ij}
\end{array}\right),\\
\quad   R_{i0j0} & = &-\Big(\dot{H}+H^2\Big)\delta_{ij}\,,\quad   R_{ijkl}=H^{2}(\delta_{ik}\delta_{jl}-\delta_{il}\delta_{jk}),\\
  \Box|_{\text{scalar}}&=&-\pd^2_t-3H\pd_t+\frac{1}{a^2}\delta^{ij}\pd_{i}\pd_{j}
\label{frwtcscurves}
\end{eqnarray}
Here $H$ is the Hubble function and dot denotes
the derivative with respect to $t$.
$\Box|_{\text{scalar}}$ is the d'Alembertian operator acting on scalars.
Small Latin letters from the middle of the alphabet
are spatial indices.  Index $t$ is used to designate a derivative w.r.t. cosmic time.

Equivalently we write the FLRW metric (\ref{frwcosmic}) as
\begin{eqnarray}
ds^{2}=a(\tau)^{2}\left(-d\tau^{2}+\delta_{ij}dx^{i}dx^{j}\right)\,.\label{frwtau}
\end{eqnarray}
$\tau$ is the conformal time such that $ad\tau=dt$. The background
quantities in the latter metric are
\begin{eqnarray}
  \Gamma_{0\nu}^{\mu} & = & \Hc\delta_{\nu}^{\mu}\,,\quad  \Gamma_{\mu\nu}^{0}={\delta_{\mu\nu}}\Hc\,,\quad\Hc=a'/a\,,\\
  R & = & \frac{6}{a^{2}}(\Hc'+\Hc^{2})\,,\quad   R_{\mu\nu}=\left(\begin{array}{cc}
-3\Hc' & 0\\
0 & (\Hc'+2\Hc^{2})\delta_{ij}
\end{array}\right)\,,\\
  G_{\mu\nu}&=&\left(\begin{array}{cc}
3\Hc^2 & 0\\
0 & -(2\Hc'+\Hc^{2})\delta_{ij}
\end{array}\right)\,,\\
  R_{i0j}^{0} & = & \Hc'\delta_{ij}\,,\quad   R_{0j0}^{i}=-\Hc'\delta_{j}^{i}\,,\quad   R_{jkm}^{i}=\Hc^{2}(\delta_{k}^{i}\delta_{jm}-\delta_{m}^{i}\delta_{kj})\,.\label{frwtaucscurves}
\end{eqnarray}
We use the index ``$0$'' for the $\tau$-component of any tensor in order not to confuse with just a small Greek letter. Prime is the derivative
with respect to the conformal time $\tau$. Further:
\begin{eqnarray}
\pd_{0} & = & a\pd_{t},~ H=\Hc/a,~  \Box|_{\text{scalar}}  =  -\frac{1}{a^{2}}(\pd_{0}^{2}+2\Hc\pd_{0}-\delta^{ij}\pd_{i}\pd_{j}).
\end{eqnarray}
Spatially flat FLRW Universe is conformally flat and the Weyl tensor for it is identically zero.

For a perfect fluid matter we assume a stress-energy tensor $T^\mu_\nu=\mathrm{diag}(-\rho,p,p,p)$ with $\rho$ to be its energy and $p$ its pressure. The equation of state parameter is defined as $w=p/\rho$. In particular, $w=1/3$ for a radiation fluid and thus it is traceless.

Weyl tensor is the totally traceless part of the Riemann tensor and is defined as follows
\begin{eqnarray}
C_{\alpha\nu\beta}^{\mu} & = & R_{\alpha\nu\beta}^{\mu}-\frac{1}{D-2}(\delta_{\nu}^{\mu}R_{\alpha\beta}-\delta_{\beta}^{\mu}R_{\alpha\nu}+{g}_{\alpha\beta}R_{\nu}^{\mu}-{g}_{\alpha\nu}R_{\beta}^{\mu})\nonumber \\
 & + & \frac{R}{(D-2)(D-1)}(\delta_{\nu}^{\mu}{g}_{\alpha\beta}-\delta_{\beta}^{\mu}{g}_{\alpha\nu})=\nonumber\\
& = & R_{\alpha\nu\beta}^{\mu}-\delta_{\nu}^{\mu}S_{\alpha\beta}+\delta_{\beta}^{\mu}S_{\alpha\nu}-g_{\alpha\beta}S_{\nu}^{\mu}+g_{\alpha\nu}S_{\beta}^{\mu}\,.
\end{eqnarray}
Here $$S_{\mu\nu}=\frac{1}{D-2}\left(R_{\mu\nu}-\frac{1}{2(D-1)}Rg_{\mu\nu}\right)$$ is the Schouten tensor.
One useful relation for the double divergence of the Weyl tensor valid in any dimension and for an arbitrary background is
\begin{equation}
\begin{split}\frac{1}{D-3}\nabla^{\alpha}\nabla^{\beta}C_{\alpha\mu\nu\beta} & =-\Box S_{\mu\nu}+\frac{1}{2(D-1)}\nabla_{\mu}\pd_{\nu}R+\frac{1}{D-2}C_{\rho\nu\mu\alpha}L^{\alpha\rho}\\
 & +\frac{D}{(D-2)^{2}}L_{\mu\alpha}L_{\nu}^{\alpha}-\frac{1}{(D-2)^{2}}g_{\mu\nu}L_{\alpha\beta}^{2}+\frac{1}{(D-1)(D-2)}RL_{\mu\nu}
\end{split}
\label{usefulDDWeyl}
\end{equation}
where $L_{\mu\nu}=R_{\mu\nu}-\frac1D R$ is the traceless part of the Ricci tensor.

We widely use the spatial Fourier transform which reads
\begin{equation}\phi(\tau,\vec x)=\int e^{i\vec k \vec x}\phi(\tau,\vec k) d\vec k.
\label{fourier3}
\end{equation}
This also allows to use a Fourier transformed  d'Alembertian
$$\Box_k|_{\text{scalar}}=-\pd_{t}^{2}-3H\pd_{t}-\frac{k^2}{a^{2}}=-\frac{1}{a^{2}}(\pd_{0}^{2}+2\Hc\pd_{0}+k^2)$$

\section{Perturbations}\label{app:perturbations}

\subsection{Covariant perturbations}
Covariant metric perturbations are introduced as
$$g_{\mu\nu}\to g_{\mu\nu}+h_{\mu\nu}$$
Covariant four dimensional perturbations of the relevant quantities then read
\begin{eqnarray}
\delta\Gamma^\mu_{\nu\rho}\equiv\gamma^\mu_{\nu\rho}&=&\frac12(\cpd_\nu h^\mu_\rho+\cpd_\rho h^\mu_\nu-\cpd^\mu
h_{\nu\rho}),~\gamma_{\mu\rho}^\rho=\frac12\pd_\mu h\\
\delta R^\sigma_{\mu\nu\rho}&=&
\cpd_\nu\gamma^\sigma_{\mu\rho}-\cpd_\rho\gamma^\sigma_{\mu\nu}\nonumber\\
&=&\frac12(\cpd_\nu\cpd_\mu h^\sigma_\rho{+[\cpd_\nu\cpd_\rho] h^\sigma_\mu}-\cpd_\nu\cpd^\sigma h_{\mu\rho}-\cpd_\rho\cpd_\mu h^\sigma_\nu+\cpd_\rho\cpd^\sigma h_{\mu\nu})
\\
\delta R_{\mu\rho}&=&\cpd_\nu\gamma^\nu_{\mu\rho}-\cpd_\rho\gamma^\nu_{\mu\nu}=\frac12(\cpd_\nu\cpd_\mu
h^\nu_\rho+\cpd_\nu\cpd_\rho h^\nu_\mu-\Box h_{\mu\rho}-\cpd_\rho\pd_\mu h)
\label{deltardd}
\\
\delta R\equiv r&=&(-R_{\mu\nu}+\cpd_\mu\cpd_\nu-g_{\mu\nu}\Box)h^{\mu\nu}
\label{deltar}
\end{eqnarray}

Given the Einstein-Hilbert action with a cosmological term $\Lambda$ such that
\begin{equation}
S_\Lambda = \int d^4x\ \sqrt{-g}\left[\frac{M_P^2}2 R-\Lambda\right],
\label{EHLambda}
\end{equation}
one can readily compute its second order variation around any background.
To this end, one has to use on top of already presented above expressions a (part of the) second variation of the Ricci tensor which can be written in terms of the variation of the Christoffel symbol as follows
\begin{equation*}
 \delta^2 R_{\mu\rho}\simeq\gamma^\sigma_{\chi\sigma}\gamma^\chi_{\mu\rho}
-\gamma^\sigma_{\chi\rho}\gamma^\chi_{\sigma\mu}+\dots
\end{equation*}
The ``part of'' means that only terms which eventually contribute to the second order variation of the latter action are written.
The resulting second variation for action (\ref{EHLambda}) is (see, e.g. \cite{Christensen:1979iy})
\begin{equation}
\begin{split}
\delta^2 S_{\Lambda}=\int
d^4x\sqrt{-  g}\frac{M_P^2}2&\left[\left(\frac14 h_{\mu\nu}  \Box
h^{\mu\nu}-\frac14h  \Box h+\frac12h  \cpd_\mu  \cpd_\rho
h^{\mu\rho}+\frac12  \cpd_\mu h^{\mu\rho}  \cpd_\nu h^\nu_\rho\right)\right.+\\
&+(hh^{\mu\nu}-2h^\mu_\sigma h^{\sigma\nu})\left(\frac18
  g_{\mu\nu}  R-\frac14   g_{\mu\nu}\frac\Lambda{M_P^2}-\frac12   R_{\mu\nu}\right)-\\
&-\left.\left(\frac12  R_{\sigma\nu}h^\sigma_\rho
h^{\nu\rho}+\frac12  R^\sigma_{\rho\nu\mu}h^\mu_\sigma h^{\nu\rho}\right)\right].
\end{split}
\label{deltaEHfull}
\end{equation}
This is a ready-to-go expression to compute a second variation of the Einstein-Hilbert action with $\Lambda$ for any line element.

In studying perturbations around MSS backgrounds York decomposition for the rank-2 tensor can be utilized to yield
\begin{equation}
  h_{\mu\nu}=h_{\mu\nu}^\perp+(\nabla_\mu A^\perp_\nu+\nabla_\nu A^\perp_\mu)+\left(\nabla_\mu\nabla_\nu -\frac1Dg_{\mu\nu}\Box\right)B+\frac1Dg_{\mu\nu}h
  \label{hmunucovariant210}
\end{equation}
This formula can be found as (5.1) in \cite{DHoker:1999bve} for instance.
$h_{\mu\nu}^\perp$ is transverse and traceless tensor having $(D+1)(D-2)/2$ DOF-s, $A^\perp_\mu$ is a transverse vector with $(D-1)$ independent components and $B$, and $h$ are scalars summing up to $D(D+1)/2$ DoF-s. $D$ DOF-s can be gauged away while some DoF-s become non-dynamical due to constraints. On MSS $h_{\mu\nu}^\perp$ is a spin-2 field, and $A^\perp_\mu$ is a spin-1 field. The real physical graviton on MSS has $D-2$ polarizations and is a subset of $h^\perp_{\mu\nu}$. Consequently, on MSS one can consider $h_{\mu\nu}^\perp$, $A_\mu^\perp$ and scalars independently as they do not mix carrying different spins. This makes this formalism useful for MSS backgrounds. Moreover, vector $A_\mu^\perp$ results in  no dynamical DOF-s on MSS.

Second order variations of action (\ref{eq:masterL}) for the trace part and for the transverse and traceless part around a MSS background can be found in \cite{Biswas:2016egy} and are as follows
\begin{equation}
\delta^2 S_{0}=\frac1{32}\int
dx^4\sqrt{-  g}\ h(3  \Box+  R)\left\{-\left(\frac{M_P^2}2+\lambda {f_0}  R\right)
+\lambda\Fc_R(  \Box)(3  \Box+  R)
\right\}h.
\label{d2properallRconstSCALARSnobars}
\end{equation}
\begin{equation}
\delta^2 S_2=\frac14\int
dx^4\sqrt{-  g}\ {h^\perp}_{\nu\mu}\left(   \Box-\frac {  R}6
	\right)\left\{\frac{M_P^2}2+\lambda {f_0}
  R+
\lambda
\Fc_C\left(  \Box+\frac{  R}3\right)
\left( \Box-\frac{  R}3\right)\right\}{h^\perp}^{\nu\mu}.
\label{d2properallRconstTENSORS}
\end{equation}
Subscripts 0 and 2 indicate the corresponding spin.
It seems extremely difficult to work out any but MSS backgrounds using the covariant formalism.
The $(1+3)$ formalism allows quite an easy extension to FLRW space-times.

\subsection{Scalar perturbations in $(1+3)$ decomposition}

We follow the standard procedure of introducing Bardeen variables. For scalars the perturbed line element is:
\begin{equation}
ds^2=a(\tau)^2\left[-(1+2\phi)d\tau^2-2\pd_i \beta d\tau
dx^i+((1-2\psi)\delta_{ij}+2\pd_i\pd_j\gamma)dx^idx^j\right],
\label{FLRWscalars}
\end{equation}
where $\tau$ is the conformal time
and $\phi,\beta,\gamma,\psi$ are 4 scalar DOF-s without any gauge fixing yet. In fact two scalars can be gauged away.
The following combinations are gauge invariant and are named Bardeen potentials
\begin{equation}
\Phi=\phi-\frac{1}{a}(a\vartheta)^\prime=\phi-\dot
\chi,\qquad\Psi=\psi+\Hc\vartheta=\psi+H\chi,
\label{GIvars}
\end{equation}
where  $\chi=a\beta+a^2\dot\gamma$, $\vartheta=\beta+\gamma'$.

We proceed assuming that all the fields are Fourier transformed using (\ref{fourier3}).
Variation of Einstein equations can be written using gauge invariant variables as follows
\begin{equation*}
\begin{split}
 \delta G^0_0&=\frac{2}{a^2}[k^2\Psi+3\Hc\Psi'+3\Hc^2\Phi+3\Hc(\Hc^2-\Hc')(\beta+\gamma')]e^{i\vec k\vec x}\\
 \delta G^0_i&=-\frac{2ik_i}{a^2}[\Psi'+\Hc\Phi+(\Hc^2-\Hc')(\beta+\gamma')]e^{i\vec k\vec x}\\
 \delta G^i_j&=-\frac{k^ik_j}{a^2}[\Psi-\Phi]e^{i\vec k\vec x}~\text{ for }~i\neq j\\
 \delta G^i_i&=\frac{1}{a^2}\Big[
 2\Psi''+(k^2-k_i^2)(\Psi-\Phi)+2\Hc(2\Psi'+\Phi')+2(2\Hc'+\Hc^2)\Phi\\
 &+2[(\Hc^2-\Hc')'+\Hc(\Hc^2-\Hc')](\beta+\gamma')\Big]e^{i\vec k\vec x}~\text{ whith no sum over }~i\\
 \end{split}
\end{equation*}

Variation of the scalar curvature is
\begin{equation}
\begin{split}
  \delta R\equiv re^{i\vec k\vec x}
  &=\left[6\Box_k\Psi-2R\Phi-\frac 2{a^2}\left[3\Hc(\Psi'+\Phi')-k^2(\Psi+\Phi)\right]+R'(\beta+\gamma')\right]e^{i\vec k \vec x}
  \end{split}
  \label{deltaRscalars}
\end{equation}

Variation of the Weyl tensor with two indexes up and two down is especially simple and has the following form
\begin{equation}
 \delta C_{\mu\alpha}^{\phantom{\mu\alpha}\nu\beta}\equiv we^{i\vec k \vec x} K_{\mu\alpha}^{\phantom{\mu\alpha}\nu\beta} 
 =\frac{\Phi+\Psi}{a^2}e^{i\vec k \vec x}K_{\mu\alpha}^{\phantom{\mu\alpha}\nu\beta}
 \label{deltaWeyl}
\end{equation}
where $K_{\nu\alpha}^{\phantom{\nu\alpha}\beta\mu}$ is a constant tensor whose components depend on $k_i$ only. Moreover it is absolutely traceless and retains all the symmetry properties of the Weyl tensor. Its explicit form is
\begin{eqnarray}
 K_{0j}^{\phantom{0j}0i}&=& -\frac16k^2\delta^i_j+\frac12k^ik_j ,\nonumber\\
K_{mj}^{\phantom{mj}ki}&=& \frac13k^2(\delta^k_m\delta^i_j-\delta^k_j\delta^i_m)
-\frac12\delta^k_mk^ik_j
-\frac12\delta^i_jk^kk_m
+\frac12\delta^k_jk^ik_m
+\frac12\delta^i_mk^kk_j .
\end{eqnarray}

Second variation of the Einstein-Hilbert action with a cosmological constant (\ref{EHLambda}) around a FLRW background can be straightforwardly computed by substituting metric perturbations introduced in (\ref{FLRWscalars}) into (\ref{deltaEHfull}). The resulting expression is
\begin{equation}
\begin{split}
\delta^2 S_{\Lambda}=\int
d\tau&\Bigg[\frac{M_P^2}2 a^2\bigg\{-6\psi'^2+6k^2\psi^2-9\Hc^2(\phi+\psi)^2-12\Hc(\phi+\psi)\psi'-4k^2\psi(\psi+\phi)+\\
&+k^2\left(-4\Hc\phi(\beta+\gamma')+4(2\Hc^2+\Hc')\psi\gamma-4\psi'(\beta+\gamma')-6\Hc^2(\phi+\psi)\gamma+3\Hc^2\beta^2\right)\\
&-(\Hc^2+2\Hc')k^4\gamma^2\bigg\}
+\frac{\Lambda}{2}a^4\left((\psi-\phi-k^2\gamma)^2-4(\psi-\phi)^2+4\phi^2-k^2\beta^2\right)\Bigg].
\end{split}
\label{deltaEHADM}
\end{equation}
While this formula can rewritten using gauge invariant potentials, it is useless until a background which solves background equations of motion is fixed.

Note that all the expressions in this Subsection are valid for any FLRW space-time.
More complicated variations for terms beyond GR in (\ref{eq:masterL}) can be obtained by combining written above variations of the scalar curvature and the Weyl tensor as long as conformally flat backgrounds are considered. All computations become excessively lengthy beyond conformally flat space-times.

\subsection{Tensor perturbations in $(1+3)$ decomposition}

For tensor perturbations the line element is:
\begin{equation}
	ds^2=a(\tau)^2\left[-d\tau^2+(\delta_{ij}+2\chi_{ij})dx^idx^j\right],
\label{mFrtensors}
\end{equation}
and the perturbations are traceless and transverse (in 3-dimensional sense), i.e. $\pd_i\chi^i_j=\chi^i_i=0$. Moreover, tensor perturbations are gauge invariant. We follow the standard convention that indexes of $\chi_{ij}$ are manipulated using the 3-dimensional metric, $\delta_{ij}$ in our case. If embedded into $h_{\mu\nu}$ as follows $h_{ij}=2a^2\chi_{ij}$ and $h_{0\nu}=0$ then the resulting $h_{\mu\nu}$ is obviously traceless in the 4-dimensional sense, i.e. $h^\mu_\mu=0$. It also can be readily shown to be also transverse in the 4-dimensional sense, i.e. $\nabla_\mu h^\mu_\nu=0$. While $h_{ij}$ is manipulated using the 4-dimensional metric, we note that $h^i_j=2\chi^i_j$. In particular this means that the variation of the scalar curvature vanishes on any FLRW background, i.e.
\begin{equation}\delta_T R=0
\end{equation}
where we use the subscript $T$ indicating that tensor perturbations are considered.
This happens because the spatial part of the Ricci tensor is proportional to $g_{ij}$ while other terms give a divergence and a trace (see (\ref{deltar})). Variation of the Einstein tensor becomes
\begin{equation}
 \delta_T G^i_j=\delta_t R^i_j=-\left(\Box-2\frac{\Hc^2}{a^2}\right)\chi^i_j
\end{equation}
with all other components zero.
The second variation of action (\ref{EHLambda}) can also be straightforwardly computed to give
\begin{equation}
 \delta_T^2 S_\Lambda=\int
d^4x\sqrt{-g}\frac{M_P^2}2\chi^i_j\left(\Box-\frac23R+\frac{2\Lambda}{M_P^2}\right)\chi^j_i
\label{delta2Stensors}
\end{equation}

The formulae in this Subsection are valid for any FLRW background.
Further expressions like variation of the Weyl tensor are becoming long and not illuminating. A second order variation of the initial action (\ref{eq:masterL}) can be formally written for tensor modes but is again very lengthy and not a priori useful.

\providecommand{\href}[2]{#2}\begingroup\raggedright\endgroup

 \end{document}